\documentclass[sigconf,9pt]{acmart}

\usepackage[english]{babel}
\usepackage{blindtext}

\usepackage{tabularx,tablefootnote,wrapfig,multirow}
\usepackage{hyphenat,xspace,color,colortbl,enumitem}
\usepackage{booktabs,multirow}
\usepackage{bm}
\usepackage[normalem]{ulem}
\usepackage{newtxmath,bm,courier,textcomp}
\usepackage{enumitem,fancyhdr,fancyref}
\usepackage{xfrac,sparklines,bigstrut,physics}
\usepackage[labelformat=simple,skip=0pt]{subcaption}
\definecolor{Gray}{gray}{0.9}
\definecolor{LightGreen}{rgb}{0.88,1,0.88}
\definecolor{LightOrange}{rgb}{1,0.85,0.8}
\definecolor{LightRed}{rgb}{1,0.80,0.80}

\newcommand{\parahead}[1]{\vspace{2pt plus 0pt minus 2pt}\noindent{\bfseries #1}}
\newcommand{\parabreak}{\vspace*{1.00ex minus 0.25ex}\noindent}

\renewcommand{\paragraph}[1]{\parahead{#1}}

\definecolor{light-blue}{rgb}{0.8,0.8,1.0}

\setitemize{itemsep=1pt,topsep=2pt,parsep=1pt,partopsep=0pt,leftmargin=1.5em}
\setenumerate{itemsep=1pt,topsep=2pt,parsep=1pt,partopsep=0pt,leftmargin=1.5em}
\setlist{itemsep=1pt,parsep=1pt}

\hypersetup{colorlinks=true, urlcolor=blue,
    citecolor=black, linkcolor=black, menucolor=black}

\fancyrefchangeprefix{\fancyreffiglabelprefix}{f}
\fancyrefchangeprefix{\fancyreftablabelprefix}{t}
\fancyrefchangeprefix{\fancyrefeqlabelprefix}{eqn}

\newcommand{\systemname}{QuAMax}
\newcommand{\systemnames}{QuAMax's}

\renewcommand\footnotetextcopyrightpermission[1]{} % removes footnote with conference info
\setcopyright{none}
\begin{document}
\pagenumbering{Roman}

\title{Leveraging Quantum Annealing for Large MIMO Processing 
in \\ Centralized Radio Access Networks}

%\titlenote{Produces the permission block, and copyright information}
%\subtitle{Extended Abstract}

\author{Minsung Kim}
\affiliation{%
 \institution{Princeton University}
}
\email{minsungk@cs.princeton.edu}

\author{Davide Venturelli}
\affiliation{%
 \institution{USRA Research Institute for Advanced Computer Science}
}
\email{DVenturelli@usra.edu}

\author{Kyle Jamieson}
\affiliation{%
 \institution{Princeton Univeristy}
}
\email{kylej@cs.princeton.edu}

\begin{abstract}
User demand for increasing amounts of wireless capacity continues to outpace supply, and so to meet this demand, significant progress has been made in new MIMO wireless physical layer techniques.  Higher-performance systems now remain impractical largely only because their algorithms are extremely computationally demanding.  For optimal performance, an amount of computation that increases at an exponential rate both with the number of users and with the data rate of each user is often required. The base station's computational capacity is thus becoming one of the key limiting factors on wireless capacity.  \emph{\systemname{}} is the first large MIMO centralized radio access network design to address this issue by leveraging quantum annealing on the problem. We have implemented \systemname{} on the 2,031 qubit D-Wave 2000Q quantum annealer, the state-of-the-art in the field.  Our experimental results evaluate that implementation on real and synthetic MIMO channel traces, showing that 10~$\mu$s of compute time on the 2000Q can enable 48 user, 48 AP antenna BPSK communication at 20 dB SNR with a bit error rate of $10^{-6}$ and a 1,500 byte frame error rate of $10^{-4}$.

%We argue that using quantum assisted technology in the future could hold the potential to overcome the computational limits of conventional computers on the number of mobile users that a MIMO base station can effectively serve. Our work paves the way for quantum hardware and software to significantly expand the performance envelope of Massive MIMO.
%\textcolor{red}{[[I suggest the following revision]] Our work establishes programming techniques and baseline performance benchmarks of this new approach, paving the way to future designs of hybrid quantum-classical systems that hold the potential to significantly expand the performance envelope of Massive MIMO in the future. }
\end{abstract}

% <concept_id>10003033.10003058.10003065</concept_id>
% <concept_desc>Networks~Wireless access points, base stations and infrastructure</concept_desc>
% <concept_id>10010583.10010786.10010813.10011726</concept_id>
% <concept_desc>Hardware~Quantum computation</concept_desc>

% \ccsdesc[500]{Networks~Wireless access points, base stations and infrastructure}
% \ccsdesc[500]{Hardware~Quantum computation}

% \keywords{Wireless Networks, Massive MIMO, Maximum Likelihood Detection, Sphere Decoder, Quantum Computing, Quantum Annealing}

% \acmYear{2019}\copyrightyear{2019}
% \setcopyright{acmcopyright}
% \acmConference[SIGCOMM '19]{SIGCOMM '19: 2019 Conference of the ACM Special Interest Group on Data Communication}{August 19--23, 2019}{Beijing, China}
% \acmBooktitle{SIGCOMM '19: 2019 Conference of the ACM Special Interest Group on Data Communication, August 19--23, 2019, Beijing, China}
% \acmPrice{15.00}
% \acmDOI{10.1145/3341302.3342072}
% \acmISBN{978-1-4503-5956-6/19/08}

\maketitle

\section{Introduction}
\label{s:intro}

A central design challenge for 
future generations of wireless networks is to meet users'
ever\hyp{}increasing demand for capacity and throughput.
Recent advances in the design of wireless networks to this end,
including the 5G efforts underway in industry and academia,
call in particular 
for the use of Large and Massive \emph{multiple input multiple output} (MIMO) antenna arrays to support
many users near a wireless access point (AP) or base station 
sharing the wireless medium at the same time.\footnote{We use the terms "access point" and "base station" interchangeably in this paper.} 

Much effort has been and is currently being 
dedicated to large and massive MIMO,
and these techniques are coming to fruition, yielding 
significant gains in network throughput.
An apropos example is Massive MIMO:\ in LTE cellular
and 802.11ac local\hyp{}area networks, up to eight antennas
are supported at the AP
\emph{spatially multiplexing} \cite{telatar-capacity99}
parallel information streams concurrently to multiple receive
antennas. The technique is also known as multi\hyp{}user MIMO (MU-MIMO) and
can be used both in the uplink and the
downlink of multi\hyp{}user MIMO networks:\ in the uplink
case, several users concurrently transmit to a multi\hyp{}antenna AP,
while over the downlink, the AP multiplexes different
information streams to a number of mobile users.

From a design standpoint, one of the most promising and cost
effective architectures
to implement 5G technologies is the centralized radio access network (C-RAN) architecture
\cite{namba2012bbu, rost2014opportunistic}. C-RAN 
pushes most of the physical\hyp{}layer processing that currently
takes place at the AP to a centralized data center, where it is aggregated with other APs' processing on the same
hardware.  The C-RAN concept has undergone several iterations 
since its inception, with more recent work treating
the unique demands of small cells
\cite{fluidnet-ton16}, centralizing most of the computation and supporting hundreds or thousands of the APs from a centralized data center.

To fully realize Massive MIMO's potential throughput gains, however,
the system
must effectively and efficiently demultiplex mutually\hyp{}interfering
information streams as they arrive. Current large MIMO designs such
as Argos \cite{argos-mobicom12}, BigStation \cite{BigStation}, 
and SAM \cite{SAM} use linear processing methods such as
zero\hyp{}forcing and minimum mean squared error (MMSE) filters. 
These methods have
the advantage of very low computational complexity, but suffer when
the MIMO channel is poorly\hyp{}conditioned \cite{Geosphere}, 
as is often the case when the
number of user antennas approaches the number of antennas at the AP
\cite{Geosphere, BigStation}.  Sphere Decoder\hyp{}based {\it maximum likelihood} (ML)
MIMO decoders/detectors \cite{Geosphere, ETH_HARD} can significantly improve
throughput over such linear filters, but suffer from increased 
computational complexity:\ compute requirements increase exponentially 
with the number of antennas \cite{SDComp1}, soon becoming prohibitive (Section~\ref{s:ml}). 

The problem of limited computational capacity stems from the 
requirement that a receiver's physical layer finish decoding
a frame before the sender requires feedback about its
decoding success or failure. For
Wi-Fi networks, \emph{e.g.}, this quantity 
is on the order of tens of $\mu$s, the spacing in
time between the data frame and its acknowledgement.
More sophisticated physical layers, such as 4G LTE, require
the receiver to give feedback in the context of incremental 
redundancy schemes, for the same reason; the processing time available is 3~ms for 4G LTE and 10~ms for WCDMA \cite{BigStation, dahlman20134g}.  As a result, most
current systems adopt linear filters, accepting
a drop in performance.

% In the context of error control coding, 
% despite the near\hyp{}Shannon performance of 
% prevalent LDPC and Turbo codes, the hardware computations carried out
% by conventional computers are limited by microprocessors' sequential dependencies. 
% The best decoding algorithms of these error correcting codes thus 
% involve several iterations to converge at a final result which
% further hampers performance, while area and time limitations\cite{albaharna1994area} limit
% the parallelization FPGAs can 
% achieve.

% Here, we argue the position
% that the design of efficient and fast computational
% structures is now overtaking wireless communication as 
% the most significant challenge for most high\hyp{}capacity
% wireless communication systems.

\parahead{New approach:\ Quantum computation in the data center}.
This paper explores the leveraging of quantum computation (QC) to
speed up the computation required for the ML MIMO decoder.
We place our ideas in the context of the QC currently already
realized in experimental hardware, and in the 
context of the dominant C\hyp{}RAN architecture.
Here we imagine a future quantum computer, co\hyp{}located with C\hyp{}RAN computational resources in a data center, connected to the APs via 
high\hyp{}speed, low\hyp{}latency fiber or millimeter\hyp{}wave links.

Optimization is one of the key applications the quantum community has 
identified as viable in the short\hyp{}term (\emph{i.e.}\ before quantum processors 
become scalable devices capable of error correction and universality).
While their potential in optimization is largely 
unproven, it is believed that it 
may be possible for \emph{Noisy Intermediate\hyp{}Scale Quantum}
(NISQ) devices to achieve 
polynomial or exponential speedups over the best known classical algorithms 
\cite{preskill2018quantum}. It is, however, important to leverage
understanding from current prototypes in 
order to inform the design of real\hyp{}world systems, since 
performance cannot be predicted or simulated 
efficiently, especially in the presence of device\hyp{}specific noise.
This is the approach we therefore advocate here.
Two main approaches have been identified for quantum optimization 
in NISQs:\ \emph{Quantum Annealing} (QA) and \emph{Quantum Approximate 
Optimization Algorithms} (QAOA). The former approach is a form of analog 
computation that has been developed theoretically in the early 
nineties \cite{finnila1994quantum}, but realized experimentally in a 
programmable device only in 2011 by D-Wave Systems.
We focus on QA here, discussing QAOA briefly in
Section~\ref{s:related}.

\parabreak{}This paper presents \emph{\systemname{}}, the 
first system to apply QA to the computationally challenging ML 
MIMO wireless decoding problem in the context of a centralized RAN
architecture where a QA is co-located in a data center serving
one or more wireless APs. The contributions of our 
paper can be summarized as follows: 
Firstly, we present the first reduction of the ML MIMO decoding
problem to a form that a QA solver can process.
Secondly, we introduce a new, communications\hyp{}specific
evaluation metric, 
\emph{Time-to-$\text{BER}$ ($\text{TTB}$)}, which evaluates 
the performance of the QA as it aims to achieve
a target bit error rate (BER) on the decoded data. Finally, 
we evaluate \systemname{} with various 
scenarios and parameter settings and test their
impact on computational performance.  
To achieve a BER of $10^{-6}$ and a frame
error rate of $10^{-4}$, ML MIMO detection on the 
D-Wave 2000Q quantum annealer requires 10--20~$\mu$s of computation time for 48\hyp{}user, 48\hyp{}AP antenna binary
modulation or 30--40~$\mu$s for $14\times 14$ QPSK at 20~dB SNR, and with the real\hyp{}world trace of $8\times8$ MIMO channel, the largest spatial multiplexing MIMO size publicly available for experiments~\cite{Argos}, \systemname{} requires 2~$\mu$s for BPSK and 2--10~$\mu$s for QPSK.
%10--600~$\mu$s for 16-QAM modulation.
% \textcolor{brown}{However, we discuss how these performance measurements are still not practical for deployment today, due to required engineering advances in system integration (see Section~\ref{s:Discussion}).}
% \newpage
\parahead{Paper roadmap.} The next section is a background primer on ML detection and QA.
Section~\ref{s:design} details our programming of the ML problem
on the QA hardware. 
Section~\ref{s:impl} describes \systemname{} implementation in further detail, 
followed by our evaluation in Section~\ref{s:experiments}. We conclude with a review of related work (Section~\ref{s:related}), discussion of current status of technology and practical considerations (Section~\ref{s:Discussion}), and final considerations (Section~\ref{s:concl}).

\section{Background}
\label{s:primer}
In this section, we present primer material on
the MIMO ML Detection problem (\S\ref{s:ml}), 
and Quantum Annealing (\S\ref{s:qa}).

\begin{table}[t!]
%\begin{table}[htbp]
\centering
\begin{small}
\caption{\normalfont  Sphere Decoder visited node 
count \cite{Geosphere}, complexity over 10,000 instances,
and practicality on a 
Skylake Core i7 architecture.}
\label{t:visited_nodes}
\begin{tabularx}
{\linewidth}{*{2}{X}cr}\toprule
{\bf BPSK}& {\bf QPSK} & {\bf 16-QAM} &{\bf Complexity (Visited Nodes)}\\ \cmidrule(r){1-4}
%$\mathbf{2\times 2}$&45&  \cellcolor{LightGreen}1.2\\
$\mathbf{12\times 12}$&$\mathbf{7\times 7}$ & $\mathbf{4\times 4}$&  \cellcolor{LightGreen} 	$\approx$ 40 (feasible)\\ %41.7
$\mathbf{21\times 21}$&$\mathbf{11\times 11}$ & $\mathbf{6\times 6}$&  \cellcolor{LightOrange} 	$\approx$ 270 (borderline) \\ %267.9
$\mathbf{30\times 30}$&$\mathbf{15\times 15}$ & $\mathbf{8\times 8}$&  \cellcolor{LightRed} 	$\approx$ 1,900 (unfeasible)\\ %1868.0 
\bottomrule
\end{tabularx}
\end{small}
\end{table}

\phantomsection
\subsection{Primer:\ Maximum Likelihood Detection}
\label{s:ml}
Suppose there are $N_t$ mobile users, each of which has one antenna, each sending
data bits to a multi\hyp{}antenna ($N_r\geq N_t$) MIMO AP based on OFDM, the dominant physical layer technique in broadband wireless communication systems~\cite{nee2000ofdm}.  Considering
all users' data bits together in a vector whose elements 
each comprise a single user's data bits, the users first map
those data bits into a complex\hyp{}valued \emph{symbol} $\mathbf{\bar{v}}$ 
that is transmitted over a radio channel:
$\mathbf{\bar{v}} = [\bar{v}_1, \bar{v}_2, \dots, \bar{v}_{N_t}]^\intercal$ $\in{\mathbb{C}^{N_t}}$.
Each user sends from a \emph{constellation} $\mathcal{O}$ of size
$|\mathcal{O}|$ = $2^Q$ ($Q$ bits per symbol).
The MIMO decoding problem,
whose optimal solution is called the ML solution,
consisting of a search over the
sets of transmitted symbols, looking for the set that
minimizes the error with respect to what has been received
at the AP:
\begin{equation}   
\label{eqn:ml}
\hat{\mathbf{v}} = \arg\min_{\mathbf{v} \in \mathcal{O}^{N_t}}
\left\lVert\mathbf{y} - \mathbf{Hv}\right\rVert^2.
\end{equation}
The ML decoder 
then ``de-maps'' the decoded symbols $\hat{\mathbf{v}}$ to 
decoded bits \label{def:decodedbit} $\hat{\mathbf{b}}$. 
In Eq.~\ref{eqn:ml}, $\mathbf{H}$ $\in{\mathbb{C}^{N_r\times N_t}} = \mathbf{H}^{I} + j\mathbf{H}^{Q}$ is the wireless channel\footnote{The channel changes every channel \emph{coherence time},
and is practically estimated and tracked via 
preambles and\fshyp{}or pilot tones. Typical coherence time
at 2~GHz and a walking speed is 
{\itshape ca.} 30~ms \cite{tse-viswanath}.} on each OFDM 
subcarrier and $\mathbf{y}$ $\in{ \mathbb{C}^{N_r}}$ ($= \mathbf{H}\mathbf{\bar{v}}+\mathbf{n}$) is the received set of symbols, perturbed by $\mathbf{n}$ $\in{\mathbb{C}^{N_r}}$, additive white Gaussian noise (AWGN)\label{def:AWGN}.
This solution minimizes detection errors, thus maximizing throughput ({\it i.e.}, throughput\hyp{}optimal decoding).

The \emph{Sphere Decoder} \cite{Agrell02,Damen03} is a
ML detector that
reduces complexity with respect to brute\hyp{}force search
by constraining its search 
to only possible sets $\mathbf{v}$
that lie within a hypersphere of radius $\sqrt{C}$ centered around $\mathbf{y}$ ({\it i.e.,} Eq.~\ref{eqn:ml} with constraint $\Vert {\textbf{y}} -\textbf{Hv} \Vert^{2}\leq C$).
It transforms Eq.~\ref{eqn:ml} into 
a tree search \cite{SD} by QR decomposition 
$\textbf{H}=\textbf{QR}$, where $\textbf{Q}$ is orthonormal and 
$\textbf{R}$ upper\hyp{}triangular, resulting in
$\hat{\textbf{v}}=\text{arg} \min_{\textbf{v}\in \mathcal{O}^{N_{t}}}   \Vert \bar{\textbf{y}} -\textbf{Rv} \Vert^{2}$, 
with $\bar{\textbf{y}}=\textbf{Q}^{*}\textbf{y}$. 
The resulting tree has a height of
$N_{t}$, branching factor of $|\mathcal{O}|$, and $1+\sum^{N_t}_{i=1} |\mathcal{O}|^i$ nodes. 
ML detection becomes  
the problem of finding the single leaf 
among $|\mathcal{O}|^{N_t}$ with minimum metric; the corresponding
tree path is the ML solution. Thus, the $\min$ in 
Eq.~\ref{eqn:ml} is a search
in an exponentially\hyp{}large space of transmitted symbols
$\left\{ \mathbf{v} \right\}$, despite Sphere Decoder reductions in
the search space size \cite{SD}. \parahead{}

% \footnote{\textcolor{red}{Each node at a certain level $l$ of the 
% Sphere Decoder search tree is associated with a partial symbol 
% vector $\mathbf{v}_l= \left[v\left(N_t-l\right), \ldots, v\left(N_t\right)\right]^T$ and is characterized by its \emph{Partial Euclidean Distance} (PD) \cite{SD}. If no child nodes satisfy the constraint $PD<C$, all the nodes originating from that parent node are discarded and the search continues from a different parent node.}} 
% for the most likely combination, which minimizes the foregoing distance calculation.

Table~\ref{t:visited_nodes} shows the average number of tree nodes visited
to perform
ML Sphere decoding, with clients transmitting modulation 
symbols on 50 subcarriers over a 20~MHz, 13~dB SNR (Signal to Noise Ratio) Rayleigh channel. The table is
parameterized on the number of clients and 
AP antennas, and the modulation, highlighting the exponential 
increase in computation. For 8 clients 
with 16\hyp{}QAM symbols, 15 clients with QPSK symbols, or 30 clients sending binary (BPSK) symbols, the Sphere Decoder visits close to 2,000 tree nodes, 
saturating, for example, Intel's Skylake core i7
architecture, whose arithmetic subsystem achieves an order of magnitude less computational throughput \cite{flexcore-nsdi17}.
Since traditional silicon's clock speed
is plateauing~\cite{elec:courtland}, the problem is especially acute.
% \begin{table}[htb]
% \centering
% \begin{small}
% \begin{tabularx}{\linewidth}{*{3}{X}}\toprule
% {\bf Size (Users $\times$ AP antennas)}& {\bf Throughput (Mbit/s)} & {\bf Complexity (GFLOPS)}\\ \cmidrule(r){1-3}
% %$\mathbf{2\times 2}$&45&  \cellcolor{LightGreen}1.2\\
% $\mathbf{4\times 4}$&100 &  \cellcolor{LightGreen}13\\
% $\mathbf{6\times 6}$&162 &  \cellcolor{LightOrange}105\\
% $\mathbf{8\times 8}$&223 &  \cellcolor{LightRed}837\\\bottomrule
% \end{tabularx}
% \end{small}
% \caption{Throughput and computation required 
% for a Sphere Decoder implementation \cite{Geosphere}.
% \normalfont 
% Cell color indicates practicality on a 
% Skylake Core i7 architecture (green ($4\times 4$): feasible, orange ($6 \times 6$): 
% borderline, red ($8\times 8$): unfeasible).}
% \label{t:sd_v_zf}
% \end{table}

% \begin{figure}
%     \centering
%     \includegraphics[width=\linewidth]{dwave_rigetti.png}
%     \caption{[From NASA Arc] D-Wave 2000Q machine 
%     installed at QuAIL.  %{\it Right:} A 19\hyp{}qubit digital quantum processor
% %manufactured by Rigetti Computing (microscope view %from \cite{otterbach2017unsupervised}).}
%     \label{f:hw:dwave_rigettipic}
% \end{figure}

\subsection{Primer:\ Quantum Annealing}
\label{s:opt}\label{s:opt:prior}\label{s:qa}

Quantum Annealers \cite{AQC, QA} are specialized,
analog computers that
solve NP\hyp{}complete and NP\hyp{}hard
optimization problems on current hardware, 
with future potential for substantial speedups
over conventional computing \cite{Mc}.  
Many NP\hyp{}hard problems can
be formulated in the Ising model \cite{10.3389/fphy.2014.00005} 
(\emph{cf.} \S\ref{s:reducing}), which
many QA machines use as input \cite{DW_part,DW_map}. NP\hyp{}complete and NP\hyp{}hard problems other than ML MIMO detection in the field of (wireless) networking that potentially benefit from QA include MIMO downlink precoding \cite{mazrouei2012vector}, channel coding \cite{jose2015analysis, wu2003block}, network routing \cite{chen2002efficient}, security \cite{forouzan2007cryptography}, and scheduling \cite{liu2001opportunistic, georgiadis2006resource}.
%On existing QA hardware, researchers have observed
%evidence of quantum tunneling and quantum entanglement 
%effects \cite{boixo-signature13}
%and have determined a potential of $10^8$ prefactor 
%speedup over conventional computation.

\begin{figure}
\centering
\includegraphics[trim={1cm 2.2cm 1cm 2cm},clip,width=0.90\linewidth]{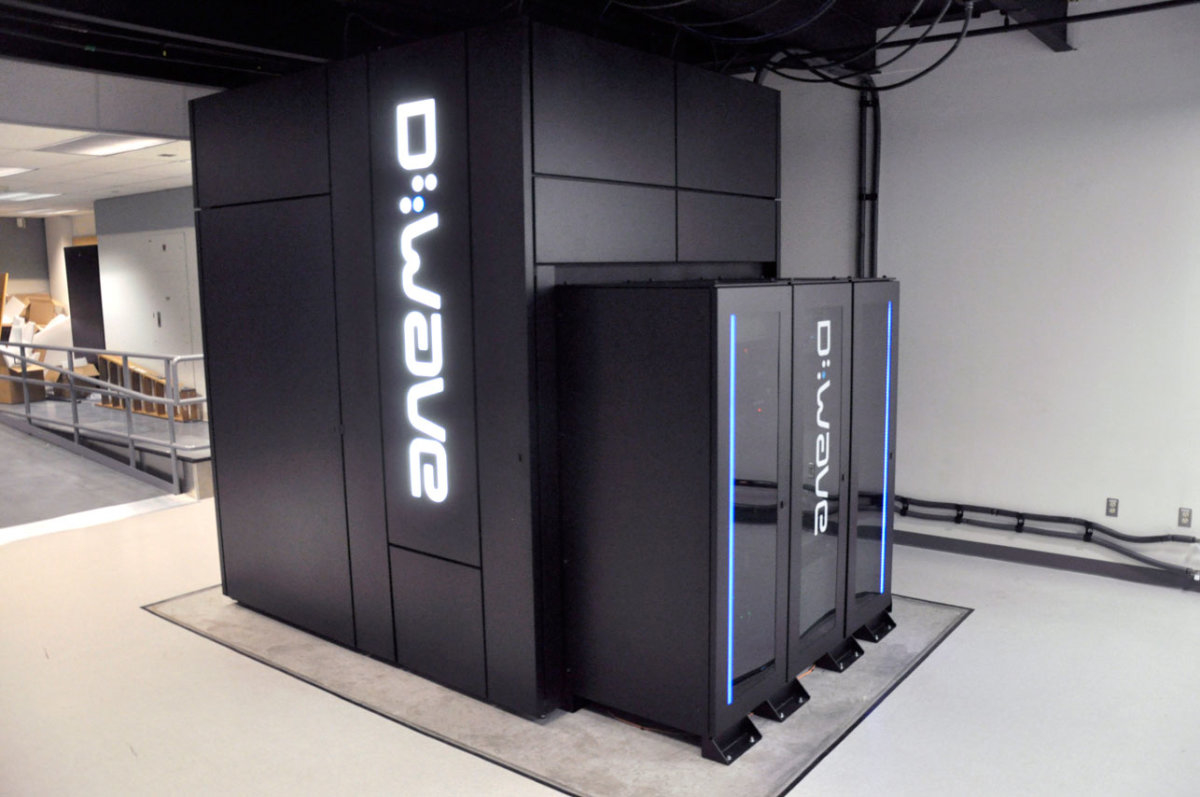}

% removed citation for anonymity:
\caption{\normalfont A D-Wave 2000Q (DW2Q) machine at NASA Ames Research Center, which
hosts a \emph{Whistler} processor manufactured with
2,048 qubits %(only 2,031 of which are calibrated, with the remainder unusable) 
and 5,019 qubit-coupling parameters. 
%(\emph{i.e.} non\hyp{}zero $Q_{ij}$ coefficients for 
%$i\neq j$ in Eq.~\ref{eqn:qubo}). 
The chip is hosted in a high\hyp{}vacuum, 
magnetically shielded enclosure at a temperature of about 13~milliKelvin.}
\label{f:dwave}
\end{figure}

\paragraph{Quantum Annealing hardware.} 
Compared to simulated annealing, the classical algorithm from which QA 
inherits its name, QA 
aims to exploit quantum effects such as \emph{tunneling},  
\emph{many-body delocalization} and \emph{quantum relaxation} to 
circumvent computational bottlenecks that would otherwise trap 
Monte Carlo methods in local minima of the solution landscape.
While exploiting QA is
technologically challenging, with the appearance 
of the D-Wave quantum annealer (Fig.~\ref{f:dwave}), the research community
is now able to run experiments, and critically, to study under 
what conditions a noisy\hyp{}intermediate\hyp{}scale\hyp{}quantum 
(NISQ) machine~\cite{preskill2018quantum} can
use quantum resources to deliver a speedup~\cite{job2018test}.
For instance, recently Boixo \emph{et al.} \cite{boixo-45467} and Denchev 
\emph{et al.} \cite{denchev-44814} have found evidence that 
tunneling under ideal conditions can be exploited on an 
earlier model of the D-Wave
2000Q (DW2Q) machine, delivering many orders of magnitude
speedup against CPU-based simulated annealing, which is considered to be one of the best classical competitions to \emph{Quantum Processing Units} (QPUs). QPUs also outperform GPU implementations by several orders of magnitude in random problems whose structure is related to real world optimization problems~\cite{king2019quantum}.

The DW2Q is an analog optimizer, meaning that it 
computes continuously rather than in discrete clock cycles,
and that it represents numerical quantities as analog instead of 
digital quantities. 
The hardware initializes each of its $N$ constituent \emph{quantum bits}, 
or \emph{qubits}, to begin in a 
\emph{superposition state} $1/\sqrt{2}
\left( \ket{0} + \ket{1} \right)$ that has no classical counterpart. In concrete terms, these qubits are metallic circuits in a chip that are maintained in a superconducting state by low temperature and subjected to the influence of tailored magnetic fluxes.
The collection of $N$ qubits at this point in time
encodes all the possible $2^N$ outputs in a single state. 
This initial setting
is achieved by exposing all the qubits in the chip to a signal 
$A(t)$ whose magnitude at this point in time is maximal. 
Then the system implements an \emph{objective function} which is 
represented by another signal $B(t)$ and is ramped up from zero, 
while $A(t)$ is decreased progressively at the same time.
The synchronized sequence of signals $A$ and $B$ and their time 
dependence is the \emph{annealing  schedule}.
The schedule is essentially the QA algorithm, and has to be optimized so 
that at the end of the run ($B(t)=\max$ and $A(t)=0$), each qubit in 
the chip assumes either a value of $\ket{0}$ or $\ket{1}$, corresponding to classical bit values, 0 or 1, respectively. This final state of these qubits collectively 
represents a candidate solution of the problem, 
ideally the \emph{ground state} of the system (\emph{i.e.}, the minimum 
of the optimization objective function) \cite{johnson-nature11, quant-ph/0001106}. 

\parabreak{}
In practice, at the end of the run, the ground state will be found with 
a probability that depends on the degree to which 
the schedule is optimal for the  problem at hand, as well as on the effect
of uncontrollable QA noise and environmental interference on the annealer. %\textcolor{brown}{Under DW2Q technological constraints, there is no general theoretical guidance on how to optimize the probability of successful runs on recent QA machines 
%such as the DW2Q to reach the ultimate limit, so one proceeds by empirical investigation \textcolor{red}{(**REDUNDANT TO BELOW ?)}}. %the user can adopt several broad mechanisms. 
While the quantum community is investigating 
physics principles to guide schedule parameters, most clearly\hyp{}understood theoretical principles do not
apply to current, imperfect experimental systems~\cite{job2018test}. Hence
the empirical approach, which we take in this paper, represents 
current state\hyp{}of\hyp{}the\hyp{}art~\cite{ronnow2014defining}. Three degrees of freedom are specifically investigated in this work.

\parabreak{}
%\begin{enumerate}
\begin{itemize}
    \item First, there are many ways of mapping a problem to an equivalent Ising formulation
    that runs on the machine (we investigate one such mapping in Section~\ref{s:design}). 
    \item Second, the user may accelerate or delay $A(t)$/$B(t)$ evolution, thus determining \emph{annealing time} (1--300~$\mu$s), the duration of 
     the machine's computation. 
    \item Finally, the user may 
     introduce stops (\emph{anneal pause})
%and \textcolor{red}{delays (\emph{anneal offsets})}
     in the annealing process, which have
been shown to improve performance in certain settings \cite{DWPauseMarshall}.
\end{itemize}
%\end{enumerate}
\section{Design}
\label{s:design}

Starting from the abstract QA problem form (\S\ref{s:qa_prob}),
\systemnames{} design reduces ML detection 
to form (\S\ref{s:reducing}), then compiles it on actual 
hardware, a process called \emph{embedding} (\S\ref{s:mapping}).

\phantomsection
\subsection{QA Problem Formulation}
\label{s:qa_prob}

The first step in leveraging QA for any problem is to define the
problem of interest as an objective function to be minimized, consisting 
of a quadratic polynomial of binary variables.
We now introduce two equivalent forms of 
this objective functions, as is customary in the QA 
application literature.

\paragraph{1. Ising spin glass form.}  In this form the
solution variables are traditionally referred to as 
\emph{spins} $s_i \in \left\{ +1, -1\right\}$.
\begin{eqnarray}
\hat{s}_1, \ldots, \hat{s}_N = 
    \arg\min_{\left\{ s_1, \ldots, s_N \right\}} \left(\sum^{N}_{i<j} g_{ij} s_i s_j + \sum^{N}_{i} f_i s_i\right)
\label{eqn:ising}
\end{eqnarray}
where $N$ is the number of spin variables,
and $g_{ij}$ and $f_{i}$ are the Ising \emph{model parameters} that 
characterize the problem.
The $f_{i}$ characterize the preference 
for each spin to be $+1$ or $-1$: positive
indicates a preference for $-1$ while negative 
indicates a preference for $+1$, with the magnitude
corresponding to the magnitude of the preference
for either state.
The $g_{ij}$ capture preferred correlations 
between spins: positive causes the QA
to prefer $s_i \neq s_j$, while negative causes
the QA to prefer $s_i = s_j$ in its optimization outcome.
Analogously to $f_i$, the magnitude of $g_{ij}$ corresponds
to the magnitude of its preference.

\paragraph{2. QUBO form.}  The 
\emph{Quadratic Unconstrained Binary Optimization} (\emph{QUBO}) 
has solution variables $q_i$ that are 
classical binary bits (zero\hyp{} or  one\hyp{}valued):
\begin{align}
\hat{q}_1,\ldots,\hat{q}_N=
\arg\min_{\left\{ q_1, \ldots, q_N \right\}} \sum^{N}_{i\leq j} Q_{ij}q_i q_j,
\label{eqn:qubo}
\end{align}
where $N$ is the qubit count and 
$\mathbf{Q} \in \mathbb{R}^{N
\times N}$ is upper triangular. The off\hyp{}diagonal matrix
elements $Q_{ij}$ ($i\neq j$) correspond
to $g_{ij}$ in Eq.~\ref{eqn:ising}, and the diagonal
elements correspond to $f_i$.

\parabreak{}The two forms are equivalent, 
their solutions related by: 
\begin{equation}
q_i \leftrightarrow \frac{1}{2}(s_i+1), 
\label{eqn:qubo2ising}
\end{equation}
%\begin{equation}
%q_i \leftrightarrow \frac{1}{2}(s_i+1), 
%\label{eqn:qubo2ising}
%\end{equation}
leading to $g_{ij} \leftrightarrow \frac{1}{4}Q_{ij}$ and 
$f_i \leftrightarrow \frac{1}{2}Q_{ii} + 
\frac{1}{4} \sum^{i-1}_{k=1} Q_{ki} + \frac{1}{4} \sum^{N}_{k=i+1} Q_{ik}$.

\subsection{ML-to-QA Problem Reduction} 
\label{s:reducing}

We now explain our process for transforming
the ML detection problem into the QUBO and Ising forms. Since \systemname{} also assumes OFDM where the wireless channel is subdivided into multiple flat\hyp{}fading orthogonal subcarriers~\cite{nee2000ofdm}, this ML\hyp{}to\hyp{}QA reduction is required at each subcarrier.

\subsubsection{ML-to-QUBO problem reduction.}
\label{s:mltoqubo}
Let's first consider the transformation 
of the ML problem into QUBO form---the key idea 
is to find a \emph{variable\hyp{}to\hyp{}symbol transform function} 
$\mathbf{T(\cdot)}$ that represents the ``candidate'' vector 
$\mathbf{v}$ in the ML search process (Eq.~\ref{eqn:ml} on
p.~\pageref{def:AWGN})
instead with a number of QUBO solution variables.
Specifically, we represent each of the $N_t$ senders' candidate symbols
${v_i} \in \mathcal{O}$ ($1 \leq i \leq N_t$), with $\log_2({|\mathcal{O}|})$ 
QUBO solution variables, naturally requiring 
$N=N_t\cdot\log_2({|\mathcal{O}|})$ QUBO variables for $N_t$ transmitters, and 
form these QUBO variables into a vector $\mathbf{q_i}$ for each sender $i$:
$\mathbf{q_i} = \left[ q_{(i-1)\cdot\log_2(|\mathcal{O}|)+1}, \ldots,\right.$ $\left. q_{i\cdot\log_2(|\mathcal{O}|)} \right]$.  
For example, $\mathbf{T}$ recasts a $2 \times 2$ QPSK ($|\mathcal{O}|=4$) problem into a QUBO problem with four
solution variables, split into two vectors $\mathbf{q_1}=[q_1\; q_2]$ 
and $\mathbf{q_2}=[q_3\; q_4]$.  In general, the transform recasts the ML problem
of Eq.~\ref{eqn:ml} into the form
\begin{equation}   
\label{eqn:encoding}
\mathbf{\hat{q}_1},\dots,\mathbf{\hat{q}_{\mathit{N_t}}} = \arg\min_{\mathbf{q_1},\dots,\mathbf{q_{\mathit{N_t}}}}
\left\lVert\mathbf{y} - \mathbf{He}\right\rVert^2,
\end{equation}
where $\mathbf{e}=[\mathbf{T(q_1)},\dots,$ $\mathbf{T(q_{\mathit{N_t}})}]^\intercal$. Then, the resulting $N_t$ vectors $\mathbf{\hat{q}_1}, \ldots, \mathbf{\hat{q}_{N_t}}$ correspond to the $N$ QUBO solution variables, ${\hat{q}_1},\dots,$ %${\hat{q}_{{N_t}\log_2({|\mathcal{O}|)}}}$
${\hat{q}_{N}}$. Continuing our $2\times 2$ QPSK example, $\mathbf{e}$ $=[\mathbf{T(q_1)}, \mathbf{T(q_2)}]^\intercal = [\mathbf{T}([q_1,q_2]),$ $\mathbf{T} ([q_3,q_4])]^\intercal$.  
Then, Eq.~\ref{eqn:encoding} results in two ML\hyp{}decoded vectors
$\mathbf{\hat{q}_{1}}$,
$\mathbf{\hat{q}_{2}}$ (noting that $\mathbf{T(\hat{q}_{1})}$, $\mathbf{T(\hat{q}_{2})}$ corresponds to the ML solution $\mathbf{\hat{v}}=[\hat{v_1},\hat{v_2}]^\intercal$ in Eq.~\ref{eqn:ml}, 
the nearest symbol vector around received $\mathbf{y}$). The decoded vectors $\mathbf{\hat{q}_{1}}$,$\mathbf{\hat{q}_{2}}$ correspond to the four decoded QUBO variables $\hat{q}_{1},\hat{q}_{2},\hat{q}_{3},\hat{q}_{4}$ in Eq.~\ref{eqn:qubo}. 
If the transmitter's bit\hyp{}to\hyp{}symbol mapping and \systemnames{} variable\hyp{}to\hyp{}symbol transform are equivalent, then the decoded $\hat{q}_{1},\hat{q}_{2},$ $\hat{q}_{3},\hat{q}_{4}$ are the 
directly de-mapped bits, $\hat{\mathbf{b}}$ from the ML solution in Eq.~\ref{eqn:ml}.  
%, which is equivalent to de-mapped symbols``de-maps'' symbols $\hat{\mathbf{v}}$ bits $\hat{\mathbf{b}}$ in \pageref{def:decodedbit}.

When transform $\mathbf{T}$ is linear
the expansion of the norm in Eq.~\ref{eqn:encoding} 
yields a quadratic polynomial objective function, since 
$q_i^2=q_i$ for any 0 or 1\hyp{}valued $q_i$.
Then the ML problem (Eq.~\ref{eqn:ml}) transforms directly
into QUBO form (Eqs.~\ref{eqn:qubo} and~\ref{eqn:encoding}).
Our task, then, is to find  
variable\hyp{}to\hyp{}symbol linear transform functions
$\mathbf{T}$ for each of BPSK, QPSK, and 16\hyp{}QAM.

\parahead{Binary modulation.}
\label{s:reducing:binary}
If the two mobile transmitters send two signals
simultaneously, each with one of two possible 
information symbols,
their transmissions can be described with a two\hyp{}vector of symbols
$\mathbf{\bar{v}} = [\bar{v}_1, \bar{v}_2]^\intercal $ $\in \left[ \left\{ \pm 1 \right\}\right.$, 
$\left.\left\{ \pm 1 \right\} \right]^\intercal$. This type of data transmission is 
called \emph{binary} modulation, of which one popular kind is
\emph{binary phase shift keying} (BPSK).
The ML problem applied to the BPSK case where symbols $v_i$ are 
represented by $v_i =$  $\mathbf{T(q_i)} = 2q_i-1$ 
thus results in a QUBO form (a detailed derivation
can be found in Appendix~\ref{s:qubo_forms}).

\begin{figure*}
\begin{subfigure}[b]{0.2\linewidth}
\begin{picture}(90,100)(0,15)
\thicklines
\put(10,58){\line(1,0){80}}
\put(50,20){\line(0,1){80}}
\multiput(20,30)(20,0){4}{\circle*{5}}
\multiput(20,50)(20,0){4}{\circle*{5}}
\multiput(20,70)(20,0){4}{\circle*{5}}
\multiput(20,90)(20,0){4}{\circle*{5}}
\put(5,80){\small 0011}\put(28,80){\small 0111}\put(52,80){\small 1011}\put(75,80){\small 1111}
\put(5,60){\small 0010}\put(28,60){\small 0110}\put(52,60){\small 1010}\put(75,60){\small 1110}
\put(5,40){\small 0001}\put(28,40){\small 0101}\put(52,40){\small 1001}\put(75,40){\small 1101}
\put(5,20){\small 0000}\put(28,20){\small 0100}\put(52,20){\small 1000}\put(75,20){\small 1100}
\end{picture}
\caption{\systemname{} transform}
\label{f:qubit:qubit}
\end{subfigure}
\hfill
\begin{subfigure}[b]{0.2\linewidth}
\begin{picture}(90,100)(0,15)
\thicklines
\put(10,58){\line(1,0){80}}
\put(50,20){\line(0,1){80}}
\multiput(20,30)(20,0){4}{\circle*{5}}
\multiput(20,50)(20,0){4}{\circle*{5}}
\multiput(20,70)(20,0){4}{\circle*{5}}
\multiput(20,90)(20,0){4}{\circle*{5}}
\put(5,80){\small 0011}\put(28,80){\small 0100}\put(52,80){\small 1011}\put(75,80){\small 1100}
\put(5,60){\small 0010}\put(28,60){\small 0101}\put(52,60){\small 1010}\put(75,60){\small 1101}
\put(5,40){\small 0001}\put(28,40){\small 0110}\put(52,40){\small 1001}\put(75,40){\small 1110}
\put(5,20){\small 0000}\put(28,20){\small 0111}\put(52,20){\small 1000}\put(75,20){\small 1111}
\end{picture}
\caption{Intermediate code}
\label{f:qubit:binary}
\end{subfigure}
\hfill
\begin{subfigure}[b]{0.30\linewidth}
\raisebox{2.5ex}{\mbox{\includegraphics[width=\linewidth]{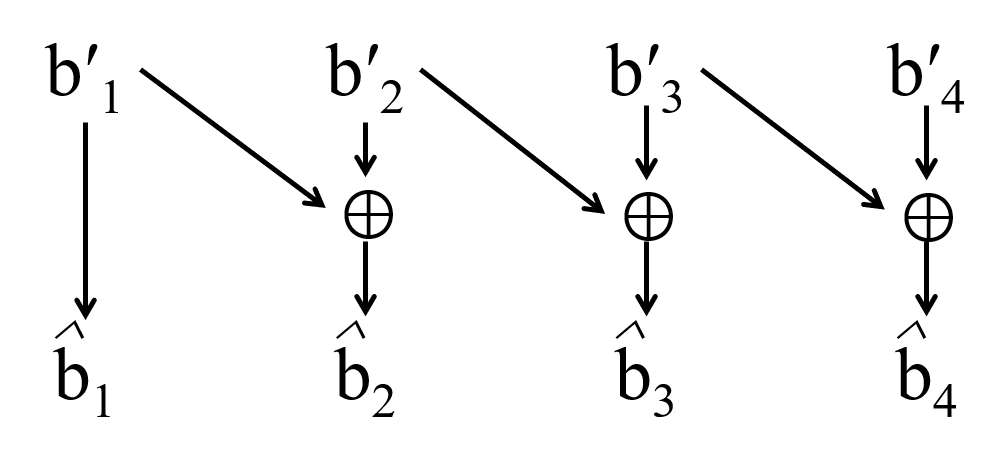}}}
\caption{Differential bit encoding}
\label{f:qubit:gray-to-binary}
\end{subfigure}
\hfill
\begin{subfigure}[b]{0.2\linewidth}
\begin{picture}(90,100)(0,15)
\thicklines
\put(10,58){\line(1,0){80}}
\put(50,20){\line(0,1){80}}
\multiput(20,30)(20,0){4}{\circle*{5}}
\multiput(20,50)(20,0){4}{\circle*{5}}
\multiput(20,70)(20,0){4}{\circle*{5}}
\multiput(20,90)(20,0){4}{\circle*{5}}
\put(5,20){\small 0000}\put(28,20){\small 0100}\put(52,20){\small 1100}\put(75,20){\small 1000}
\put(5,40){\small 0001}\put(28,40){\small 0101}\put(52,40){\small 1101}\put(75,40){\small 1001}
\put(5,60){\small 0011}\put(28,60){\small 0111}\put(52,60){\small 1111}\put(75,60){\small 1011}
\put(5,80){\small 0010}\put(28,80){\small 0110}\put(52,80){\small 1110}\put(75,80){\small 1010}
\end{picture}
\caption{Gray code}
\label{f:qubit:gray}
\end{subfigure}

\caption{\normalfont \systemnames{} bitwise post\hyp{}translation
for 16\hyp{}QAM (64\hyp{}QAM and higher\hyp{}order modulations
follow an analogous translation).}
\label{f:qubit-bit-coding}
\end{figure*}
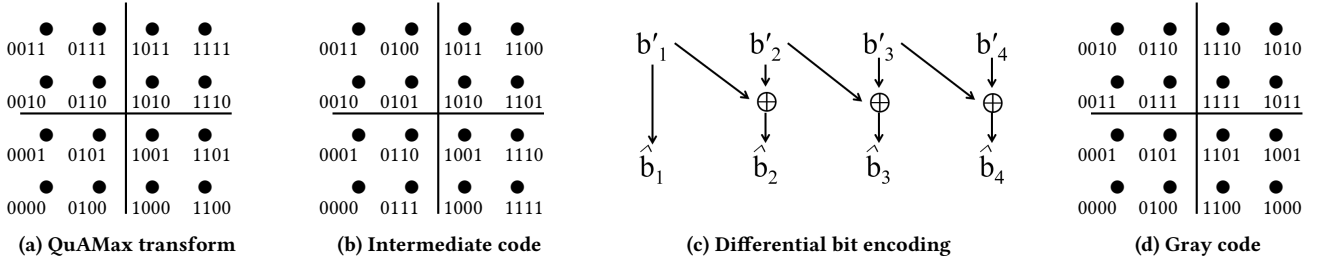

We next consider higher\hyp{}order modulations, 
which send one of $M$ possible
information symbols with each channel use (where $M>2$),
resulting in higher communication rates.
%\textcolor{red}{For these modulations,
% the sender typically applies a linear scaling to the 
% constellation to maintain a constant average transmitted
% symbol power across different order modulations}.

\parahead{QPSK modulation.}
\label{s:reducing:higher}
%\paragraph{QPSK modulation.}
In the case of \emph{quadrature phase shift
keying} (QPSK), each sender transmits one of four possible 
symbols $\bar{v}_i $ $\in \left\{ \pm 1 \pm 1j \right\}$.  Since it can be viewed as a two\hyp{}dimensional BPSK $v_i= v_i^{I} + j v_i^{Q}$, we 
represent each possibly\hyp{}transmitted QPSK information symbol with 
the linear combination of one QUBO variable, plus the other
QUBO variable times the imaginary unit. 
Transforming $q_{2i-1}$ and $q_{2i}$ to $v_i^{I}$ and $v_i^{Q}$ 
respectively leads to the transform 
$v_i = \mathbf{T(q_i)} = (2q_{2i-1}-1) + j (2q_{2i}-1)$. 

\parahead{Higher-order modulation.}
16 \emph{quadrature amplitude modulation} (16\hyp{}QAM) and higher\hyp{}order
modulations increase
spectral efficiency, but utilize multiple
amplitudes (levels) so 
require a $\mathbf{T}$ that inputs more than one 
(binary) solution variable per I or Q dimension.
First consider a
transform $\mathbf{T}$ for the simplest
multi\hyp{}level 1\hyp{}D constellation:
\mbox{\begin{picture}(80,15)
\thicklines
\put(0,0){\line(1,0){80}}
\multiput(10,0)(20,0){4}{\circle*{5}}
\put(5,5){00}
\put(25,5){01}
\put(45,5){10}
\put(65,5){11}
\end{picture}}.
$\mathbf{T} = 4q_{1}+2q_{2}-3$ maps these
bits to the values $-3,-1,+1,+3$.
Now to generalize this to 2\hyp{}D, 
let the first two arguments of 
$\mathbf{T}$, $q_{4i-3}, q_{4i-2}$, represent the
I part and the next two, $q_{4i-1}, q_{4i}$ 
represent the Q part. We call this transform,
shown in Fig.~\ref{f:qubit:qubit},
the 16\hyp{}QAM \emph{\systemname} \emph{transform}. 
It has the desirable property 
that it maps solution variables to symbols linearly, \emph{viz.}
$v_i = \mathbf{T(q_i)} = (4q_{4i-3}+2q_{4i-2}-3) + j(4q_{4i-1}+2q_{4i}-3)$, 
thus results in a QUBO form.

However, transmitters in 
practical wireless communication systems
use a different bit\hyp{}to\hyp{}symbol 
mapping, the \emph{Gray code} shown in Fig.~\ref{f:qubit:gray}, which minimizes bit errors. 
This means that the \systemname{} receiver's bit to symbol
mapping differs from the sender's. Thus one further step remains 
so that we may map the decoded QUBO variables into the 
correct Gray\hyp{}coded transmitted bits.
%(\emph{e.g.} {\tt 1100} 
%in Figure~\ref{f:qubit:qubit} 
%maps to {\tt 1000} in Figure~\ref{f:qubit:gray} 
%since they both correspond to the same symbol, $3-3j$).

A na\"{\i}ve approach is simply for \systemname{} to use the Gray\hyp{}coded bit\hyp{}to\hyp{}symbol mapping as its transform $\mathbf{T}$. The Gray\hyp{}coded mapping results in a
one\hyp{}dimensional \emph{4-PAM} constellation 
\mbox{\begin{picture}(80,15)
\thicklines
\put(0,0){\line(1,0){80}}
\multiput(10,0)(20,0){4}{\circle*{5}}
\put(5,5){00}
\put(25,5){01}
\put(45,5){11}
\put(65,5){10}
\end{picture}} assuming bits $00$, $01$,
$11$, and $10$ are transformed to $-3$, $-1$, $+1$, and $+3$ without loss of generality. The transform
$v_i^I = 2(2q_{4i-3}-1) + 2(q_{4i-3}-q_{4i-2})^2 -1$
%\label{eqn:transform2}
would map between a 4-PAM symbol $v_{i}^I$ and two QUBO variables $q_{4i-3},q_{4i-2}$, but the
resulting expansion of the ML norm would yield cubic
and quartic terms $q_rq_kq_l(q_p)$ for $r \neq k \neq l (\neq p)$, 
requiring quadratization with additional variables to 
represent the problem in QUBO form \cite{ishikawa2009higher, boros2014quadratization}. 

\parabreak{}Instead, we retain Gray coding at the transmitter and the \systemname{} transform
at the receiver. To correct the disparity,
we develop a \emph{bitwise post\hyp{}translation} that 
operates on \systemname{}\hyp{}transformed solution 
output bits at the receiver,
translating them back into 
Gray\hyp{}coded bits (\emph{i.e.}, moving from Fig.~\ref{f:qubit:qubit} to Fig.~\ref{f:qubit:gray}).
Starting with the \systemname{} transform shown in Figure~\ref{f:qubit:qubit}, 
if the second bit $\hat{q}_{4i-2}$ of the QUBO solution 
bits $\hat{q}_{4i-3}$, $\hat{q}_{4i-2}$, $\hat{q}_{4i-1}$, $\hat{q}_{4i}$ is {\tt 1}, 
then the translation flips
the third bit $\hat{q}_{4i-1}$ and the fourth bit $\hat{q}_{4i}$ ({\it e.g.} {\tt 1100} 
to {\tt 1111}), otherwise it does nothing.  This can be generalized to 
$2^{2n}$\hyp{}QAM ($n\geq2$) as an operation that flips even numbered
columns in the constellation upside down. 
We term the result $b'$ an \emph{intermediate code}, shown in 
Figure~\ref{f:qubit:binary}. Next, we apply the differential bit 
encoding transformation of Figure~\ref{f:qubit:gray-to-binary} to the 
intermediate code ${b}'$ to obtain the Gray\hyp{}coded bits ${\hat{b}}$ in 
Figure~\ref{f:qubit:gray} ({\it e.g.} translating {\tt 1111} to 
{\tt 1000}).

%\footnote{Our post\hyp{}translation can be generalized
%in the straightforward way for any $2^{2n}$\hyp{}QAM ($n\geq2$) modulation.}
%\begin{figure}[tb]
%\centering
%\includegraphics[width=\linewidth]{figures/qubit-precoding}
%\caption{A preliminary system block design for the qubit precoding
%process, in the modulation\fshyp{}detection context.}
%\label{f:qubit-precoding}
%\end{figure}

\paragraph{\systemname{} decoding example.}  To clarify processing across all stages, 
here we present a complete \systemname{} decoding example\label{def:decoding_example}. Suppose a client maps 
a bit string $b_{1},b_{2},b_3,b_4$ onto $\bar{v}_{1}$, one of 
the Gray\hyp{}coded 16\hyp{}QAM symbols in Figure~\ref{f:qubit:gray},
and sends $\mathbf{\bar{v}} = [\bar{v}_{1}]$ to an AP through wireless 
channel $\mathbf{H}$.  The AP receives $\mathbf{y}= \mathbf{H}\mathbf{\bar{v}} + \mathbf{n}$,
the transmitted signal perturbed by AWGN.
The steps of \systemnames{} decoding are:
\begin{enumerate}
\item Form the ML QUBO equation using $\mathbf{H}$, $\mathbf{y}$, and  $\mathbf{v} = [v_{1}] =[\mathbf{T({q_1})}]$, where $\mathbf{T(q_1)} = (4q_{1}+2q_{2}-3) + j(4q_{3}+2q_{4}-3)$, 
a linear transform based on the \systemname{} transform in Figure~\ref{f:qubit:qubit}.  

\item Solve the QUBO form of the ML detection problem on the QA machine, resulting an 
ML\hyp{}decoded vector $\mathbf{\hat{q}_1}$, comprised of
QUBO variables $\hat{q}_{1},\hat{q}_2,\hat{q}_3,\hat{q}_4$. 

\item Apply the above bitwise translation from the decoded QUBO solution output $\hat{q}_{1},\hat{q}_2,\hat{q}_3,\hat{q}_4$ to Gray\hyp{}coded received bits $ \hat{b}_{1},\hat{b}_2,\hat{b}_3,\hat{b}_4$ (from Figure~\ref{f:qubit:qubit} 
to Figure~\ref{f:qubit:gray}).
\end{enumerate}

If $\hat{b}_{1},\hat{b}_2,\hat{b}_3,\hat{b}_4 = b_{1},b_2,b_3,b_4$, 
decoding is done successfully, noting that in the case of a symbol error, 
we preserve the aforementioned advantage of Gray coding.

\subsubsection{ML-to-Ising problem reduction.}
\label{s:mltoising}

The Ising spin glass form of the ML problem can be obtained by 
%either 
simply 
transforming the resulting QUBO form (\S\ref{s:mltoqubo}) into the Ising form 
by Eq.~\ref{eqn:qubo2ising}.
%\sout{, or following the same variable\hyp{}to\hyp{}symbol 
%and expansion steps as QUBO but with different vectors and variables ($\mathbf{s_i}$, $s_i$ instead of $\mathbf{q_i}$, $q_i$) and different variable\hyp{}to\hyp{}symbol transform functions $\mathbf{T}$, where  $v_i = \mathbf{T(s_i)}$ = $s_{i}$ for BPSK, $\mathbf{T(s_i)}=(s_{2i-1}) + j (s_{2i})$ for QPSK, and $\mathbf{T(s_i)}=(2s_{4i-3}+s_{4i-2}) + j(2s_{4i-1}+s_{4i})$ for 16\hyp{}QAM. The expansion of the ML norm with this setting also brings out a quadratic polynomial objective function with spin variables ($s_i^2 = 1$), which is the Ising spin glass form.
%Ising and QUBO are equivalent forms, but the former 
%is more directly related to the hardware since 
%its parameters are actually used as programmable inputs in current 
%QA machines. In practice, \systemname{} directly exploits the resulting Ising model parameters without processing transform $\mathbf{T}$ and expansion of the norm}. \textcolor{red}{[[I don't understand - I think we are always saying the same thing. There is just ONE mapping to Ising - the final coefficients are the same. So we don't need to explain much, I think this part is redundant and confusing; let's just go on the next paragraph where we show the formula. It is quite obvious from previous explanation how they are obtained.]]} 
Due to the fact that DW2Q implements an Ising model, \systemname{} works 
by using the following generalized Ising model parameters:
%\textcolor{red}{for any $N_t$ }.
%We present the resulting Ising model parameters $f_{i}$ and $g_{ij}$ for BPSK and QPSK.

% \begin{align}
%     v_i \leftrightarrow \mathbf{T(s_i)} =
% \begin{cases}
%      2s_{i},& \text{BPSK} \notag\\
%      (2s_{2i-1}-1) + j (2s_{2i}-1), & \text{QPSK}
%      \notag\\
%      (2s_{4i-3}+s_{4i-2}-3) + j(2q_{4i-1}+s_{4i}-3), & \text{16\hyp{}QAM}
% \end{cases}\\
% \end{align}

% \begin{table}[htb]
% \centering
% \begin{tabular}{*{3}{l}}\toprule
% {\bf Antennas}& {\bf Throughput} & {\bf Complexity}\\ \cmidrule(r){1-3}
% $\mathbf{2\times 2}$&45  Mbit/s&  1.2 GFLOPS\\
% $\mathbf{4\times 4}$&100 &  13\\
% $\mathbf{6\times 6}$&162 &  105\\
% $\mathbf{8\times 8}$&223 &  837\\\bottomrule
% \end{tabular}
% \caption{A summary comparison of the throughput achieved and 
% computational rate required 
% for a Sphere Decoder implementation \cite{Geosphere}.  In practical
% 5G wireless systems, even more users and antennas are desirable.}
% \label{t:sd_v_zf}
% \end{table}

\parahead{BPSK modulation.}
Given a channel matrix $\mathbf{H}$ and vector of received signals $\mathbf{y}$, we obtain the following Ising model
parameters: %, which the QA uses to solve for the ML solution:
% \begin{align}
% f_i(\mathbf{H}, \mathbf{y}) &= -2 \langle \mathbf{H}^{I}_{(:,i)}, \mathbf{y}^{I}_{i}\mathbf{1}_{N} \rangle -2 \langle \mathbf{H}^{Q}_{(:,i)}, \mathbf{y}^{Q}_{i}\mathbf{1}_{N}\rangle \notag\\
% g_{ij}(\mathbf{H})&= 2\langle \mathbf{H}^{I}_{(:,i)}, \mathbf{H}^{I}_{(:,j)}\rangle + 2 \langle \mathbf{H}^{Q}_{(:,i)}, \mathbf{H}^{Q}_{(:,j)}\rangle
% \label{eqn:ising_coefficient}
% \end{align}
\begin{align}
f_i(\mathbf{H}, \mathbf{y}) &= -2 \left(\mathbf{H}^{I}_{(:,i)}\cdot  \mathbf{y}^{I}\right) -2 \left(\mathbf{H}^{Q}_{(:,i)}\cdot \mathbf{y}^{Q}\right), \notag\\
g_{ij}(\mathbf{H})&= 2\left(\mathbf{H}^{I}_{(:,i)}\cdot \mathbf{H}^{I}_{(:,j)}\right) + 2 \left( \mathbf{H}^{Q}_{(:,i)}\cdot \mathbf{H}^{Q}_{(:,j)}\right),
\label{eqn:ising_coefficient}
\end{align}
%\textcolor{red}{(bracket inner product might have to be replaced with dot, since in Quantum M bracket is used to represent qubit.)}
where $\mathbf{H}_{(:,i)}$ 
denotes the $i^{th}$ column of channel matrix $\mathbf{H}$.

\parahead{QPSK modulation.}
% Transforming $q_{2i-1}$ and $q_{2i}$ to $v_i^{I}$ and $v_i^{Q}$ respectively leads to the  following transform $\mathbf{T(q)}$ for QPSK, $v_i \leftrightarrow \mathbf{T(q_i)} = (2q_{2i-1}-1) + j (2q_{2i}-1)$,  %\textcolor{red}{The problem thus reduces to two BPSK problems}, one to decide the
%real part of the transmitted symbol, and the other to decide the imaginary part.  Note that both sub\hyp{}problems are
%formulated in terms of real\hyp{}valued variables in the Ising
%model, as for BPSK. 
In the case of QPSK, the following is the resulting Ising parameter $f_{i}$ for QPSK:
\begin{align}
    f_i(\mathbf{H}, \mathbf{y}) = 
\begin{cases}
    \text{if } i = 2n,\\
     -2\left(\mathbf{H}^{I}_{(:,i/2)}\cdot  \mathbf{y}^{Q}\right) +2 \left(\mathbf{H}^{Q}_{(:,i/2)}\cdot \mathbf{y}^{I}\right),\\%& \text{if } i = 2n
     \text{otherwise, }\\
     -2\left(\mathbf{H}^{I}_{(:,\lceil i/2\rceil )}\cdot  \mathbf{y}^{I}\right) -2 \left(\mathbf{H}^{Q}_{(:,\lceil i/2\rceil)}\cdot \mathbf{y}^{Q}\right). %\text{otherwise.}
\end{cases}
\label{eqn:qpsk_ising_f}
\end{align}
Since the real and imaginary terms of each symbol are independent, the coupler strength between $s_{2n-1}$ and $s_{2n}$ (or $q_{2n-1}$ and $q_{2n}$) is 0. %$i=2n-1$ and $j = i+1$ 
For other $s_i$ and $s_j$, the Ising coupler strength for QPSK is:
\begin{align}
g_{ij}(\mathbf{H}) =
\begin{cases}
     \text{if } i+j = 2n,\notag\\
     2\left(\mathbf{H}^{I}_{(:,\lceil i/2 \rceil)}\cdot  \mathbf{H}^{I}_{(:,\lceil j/2\rceil)}\right) +2 \left(\mathbf{H}^{Q}_{(:,\lceil i/2 \rceil)}\cdot \mathbf{H}^{Q}_{(:,\lceil j/2 \rceil)}\right), \notag\\ 
     \text{otherwise,} \notag\\
     \pm2\left(\mathbf{H}^{I}_{(:,\lceil i/2\rceil )}\cdot  \mathbf{H}^{Q}_{(:,\lceil j/2 \rceil)}\right) \mp2 \left(\mathbf{H}^{I}_{(:,\lceil j/2\rceil)}\cdot \mathbf{H}^{Q}_{(:,\lceil i/2 \rceil)}\right),
\end{cases} \\ 
\label{eqn:QPSK_ising_coefficient}
\end{align}
where $i < j$ and the sign of the latter case of Eq.~\ref{eqn:QPSK_ising_coefficient} is determined by whether $i = 2n$ (when $i = 2n$, then `$+$' and `$-$').

\parahead{16\hyp{}QAM modulation.} Ising parameters follow the same
structure as BPSK and QPSK and can be found 
in Appendix~\ref{s:16qm_model_parameter}.

In summary, the process to obtain the Ising spin glass form can be simplified with these generalized Ising model parameters; a \systemname{} system simply inserts the given channel $\mathbf{H}$ and received signal $\mathbf{y}$ at the receiver into these generalized forms accordingly, not requiring any computationally expensive operations ({\it i.e.} directly considering the expansion of the norm in Eq.~\ref{eqn:encoding}). %and then programs the resulting Ising model parameters on hardware. 
%the device will measure the given channel $\mathbf{H}$ and received signal $\mathbf{y}$ at the receiver and program the Ising model parameters accordingly, not requiring any computationally expensive operation ({\it i.e.} directly considering the expansion of the norm in Eq.~\ref{eqn:encoding}). 
%in the first step in previous \systemname{} decoding example on p.~\pageref{def:decoding_example}
Thus, computational time and resources required for ML-to-QA problem conversion are insignificant and can be neglected.

\begin{table}[htbp]
 \centering
 \caption{\normalfont Logical (physical) number of qubits required for various
 configurations of the
 elementary adiabatic quantum ML decoder.  For each configuration,
 {\bf bold font} indicates non-feasibility on the current (2,031 physical qubit) 
 D-Wave machine with Chimera connectivity.}
 
 %{\linewidth}{*{3}{l} X}\toprule
 \begin{small}
 %\scriptsize
 %\tiny
  \begin{tabularx}%{\linewidth}{X*{4}{c}}
  {\linewidth}{*{5}{X}}
  \toprule
  %\multicolumn{1}{|c|}{\multirow{2}[2]{*}{\textbf{Antennas}}} & \multicolumn{1}{c|}{\multirow{2}[2]{*}{\textbf{BPSK}}} & \multicolumn{1}{c|}{\multirow{2}[2]{*}{\textbf{QPSK}}} & \multicolumn{1}{c|}{\multirow{2}[2]{*}{\textbf{16-QAM}}} & \multicolumn{1}{c|}{\multirow{2}[2]{*}{\textbf{64-QAM}}}& \multicolumn{1}{c|}{\multirow{2}[2]{*}{\textbf{256-QAM}}} \bigstrut[t]\\
  %\multicolumn{1}{|c|}{} & \multicolumn{1}{c|}{} & \multicolumn{1}{c|}{} & \multicolumn{1}{c|}{} & \multicolumn{1}{c|}{}& \multicolumn{1}{c|}{} \bigstrut[b]\\
  {\bf Config.} & 
  {\bf BPSK} &
  {\bf QPSK} & 
  {\bf 16-QAM} & 
  {\bf 64-QAM}\\ %\multicolumn{2}{c}{{\bf 256-QAM}}\\
  \midrule
  %$\mathbf{2\times 2}$&
  %    \cellcolor{LightGreen}2 (4)&  
  %    \cellcolor{LightGreen}4 (8)& 
  %    \cellcolor{LightGreen}8 (24)& 
  %    \cellcolor{LightGreen}12 (48)\\ 
     % 16~log.& \cellcolor{LightGreen}80~phy.\\
 %  $\mathbf{3\times 3}$& 
 %     3& \cellcolor{LightGreen}6  &  
 %     6& \cellcolor{LightGreen}18  & 
 %     12& \cellcolor{LightGreen}48   & 
 %     18& \cellcolor{LightGreen}108 & 
 %     24& \cellcolor{LightGreen}108 \\
  $\mathbf{10\times 10}$& 
      \cellcolor{LightGreen}$10$ $(40)$ &  
      \cellcolor{LightGreen}$20$ $(120)$ & 
      \cellcolor{LightGreen}$40$ $(440)$ & 
      \cellcolor{LightGreen}$60$ $(1K)$ \\ 
     % 32& \cellcolor{LightGreen}288 \\
%   $\mathbf{5\times 5}$& 
%      5& \cellcolor{LightGreen}15 &  
%      10& \cellcolor{LightGreen}40  & 
%      20& \cellcolor{LightGreen}120   & 
%      30& \cellcolor{LightGreen}270 & 
%      40& \cellcolor{LightGreen}440 \\
  $\mathbf{20\times 20}$& 
      \cellcolor{LightGreen}$20$ $(120)$ & 
      \cellcolor{LightGreen}$40$ $(440)$ & 
      \cellcolor{LightGreen}$80$ $(2K)$ & 
      \cellcolor{LightRed}$\bf{120}$ $\bf{(4K)}$ \\
     % 80& \cellcolor{LightRed}1,680 ($\times$) \\
  $\mathbf{40\times 40}$&
      \cellcolor{LightGreen}$40$ $(440)$ &  
      \cellcolor{LightGreen}$80$ $(2K)$ & 
      \cellcolor{LightRed}$\bf{160}$ $\bf{(7K)}$ &
      \cellcolor{LightRed}$\bf{240}$ $\bf{(15K)}$ \\
      %160& \cellcolor{LightRed}6,560 ($\times$) \\
  $\mathbf{60\times 60}$& 
      \cellcolor{LightGreen}$60$ $(1K)$ &  
      \cellcolor{LightRed}$\bf{120}$ $\bf{(4K)}$ &
      \cellcolor{LightRed}$\bf{240}$ $\bf{(15K)}$ &
      \cellcolor{LightRed}$\bf{360}$ $\bf{(33K)}$ \\ 
      %400& \cellcolor{LightRed}40,400 ($\times$) \\
  \bottomrule
  \end{tabularx}
  \end{small}
 \label{tab:qubits}
\end{table}

\subsection{Embedding into QA hardware}
\label{s:hardware}\label{s:mapping}
\label{s:embedding}

\begin{figure}
%\vspace*{4ex}
    \centering
    \begin{subfigure}[b]{0.55\linewidth}
    \centering
    %\raisebox{1ex}{\mbox{\includegraphics[width=\linewidth]{figures/DWH2.png}}}
    \begin{picture}(240,120)
    \put(20,0){\includegraphics[width=120\unitlength]{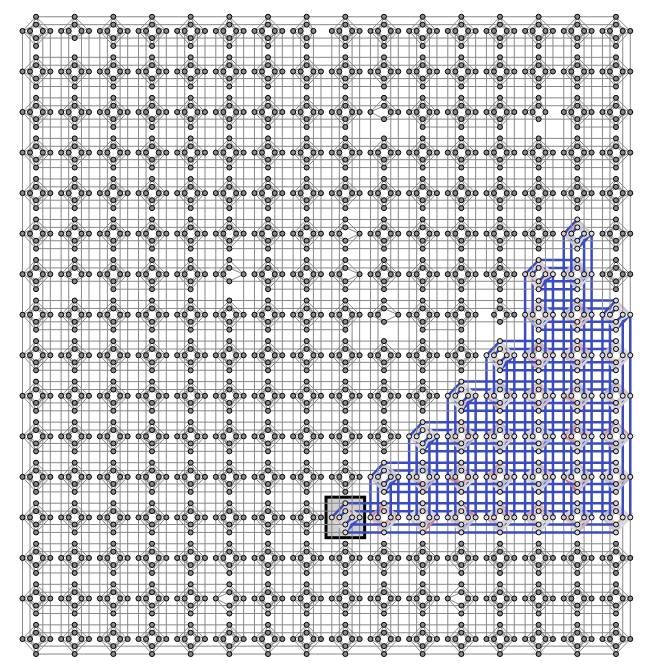}}
    \put(5,40){\includegraphics[width=80\unitlength,trim=320 120 160 400,clip]{figures/DWH2_new.png}}
    \end{picture}
    \caption{\normalfont DW2Q qubit connections:\ A  $32\times 32$ BPSK problem is shown embedded in the chip's substrate.}
    \label{f:hw:dwave}
    \end{subfigure}
    \qquad
    \begin{subfigure}[b]{0.35\linewidth}
    \includegraphics[width=\linewidth]{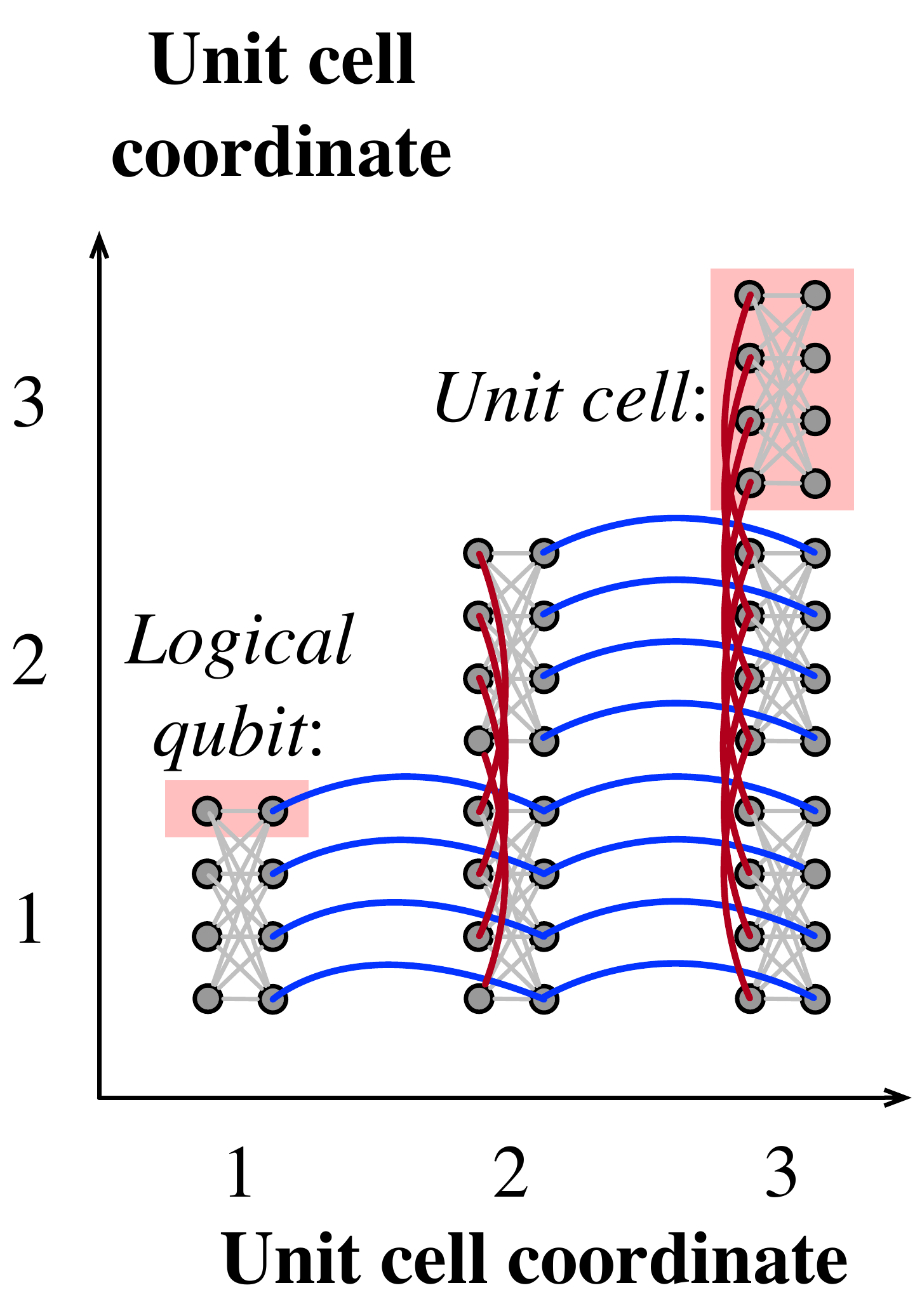}
    \caption{\normalfont Logical qubits and unit cells in the \systemname{} decoder.}
    \label{f:hw:trimap}
    \end{subfigure}
    \caption{\normalfont A comparison between the quantum hardware graph of the used machine 
    (which misses some nodes
    due to manufacturing defect), and the topology 
    of our elementary quantum ML hardware graph before embedding into the hardware graph.}
\label{f:hw}
\end{figure}

Once the ML detection problem is in
quadratic form, we still have to compile the corresponding
Ising model onto actual
QA hardware.  The D-Wave machine works by implementing 
an Ising model objective function energetically hardcoded into the chip, 
so the problem (Eq.~\ref{eqn:ising} on p.~\pageref{eqn:ising}) can 
support a certain coefficient $g_{ij}$ to be non-zero only if variables 
$s_i$ and $s_j$ are associated to physical 
variables (\emph{qubits} or \emph{physical qubits}) 
located on the chip in such a way that the qubits
are energetically coupled. In the case of the DW2Q machine we use
%\textcolor{red}{hosted at ARC} XXX TODO
the coupling matrix is a \emph{Chimera graph}, shown in 
Figure~\ref{f:hw:dwave}, with each node corresponding 
to a qubit.  Once Ising coefficients
are passed to the annealer, the hardware assigns them to 
the edges
of the Chimera graph, which are divided 
(along with their connected nodes) into \emph{unit
cells}.  Note however that, while the Ising problem generated from
Eq.~{\ref{eqn:ml}} 
is almost \emph{fully connected} (\emph{i.e.}, $g_{ij}\neq 0$ 
for most $(i,j)$ pairs), 
the Chimera graph itself has far from full connectivity,
and so a process of \emph{embedding} the Ising problem
into the Chimera graph is required.

% \begin{figure}
% \centering
% \includegraphics[width=1.0\linewidth]{figures/our_qubo.pdf}
% \caption{NOT FINAL. QPSK JF. \normalfont abc.}
% \label{f:errorbit_dist}
% \end{figure}

One standard method of
embedding is to ``clone'' variables
in such a way that a binary variable becomes 
associated not to a single 
qubit but to a connected linear chain of qubits 
instead:\ a \emph{logical qubit}, as shown in Figure~\ref{f:hw:trimap}.\footnote{The optimal 
assignment problem, in the general case, is equivalent to the NP-Hard 
``minor embedding'' problem of graph theory \cite{Choi2011}, however 
for fully\hyp{}connected graphs very efficient embeddings are 
known \cite{klymko2014adiabatic, PhysRevX.5.031040, boothby2016fast}.} We show 
an embedding of a fully-connected graph of 12~nodes. Each unit cell on the
diagonal holds four logical qubits (a chain of two qubits), while the other
unit cells are employed in order to inter\hyp{}connect
the diagonal cells. Specifically, suppose
unit cell
$[1,1]$ includes logical qubits 1--4 and unit cell
$[2,2]$ includes logical qubits 5--8. The left side
of unit cell $[2,1]$ has a vertical clone of qubits
5--8 and the right side has a horizontal clone of logical qubits
1--4. Then, logical qubits 1--4 and 5--8 are all 
connected by means of the single unit cell $[2,1]$.
The unit cell hosting the next four logical qubits 9--12 is placed 
at coordinates $[3,3]$. Two unit cells below, $[3,1]$ and $[3,2]$,
are used for connections between 9--12 and 1--4,
and 9--12 and 5--8 respectively. 
Given a number $N$ of spin variables (\emph{i.e.}, logical qubits) in Ising form, this
embedding represents each with a chain of 
$\left\lceil N/4 \right\rceil+1$  qubits, for a total 
of $N\left(\left\lceil N/4 \right\rceil+1\right)$ qubits. Recall that $N=N_t\cdot\log_2({|\mathcal{O}|})$.

% For further details on the embedding step, see Venturelli \emph{et al.}
% \cite{PhysRevX.5.031040}.\footnote{Efficient embeddings which do 
% not force the chip coverage to be a triangle are also 
% known~\cite{boothby2016fast}.}

Table~\ref{tab:qubits} summarizes the 
size of the embedding in both logical and physical qubits, as a function
of the MIMO detection problem's parameters---number of users and AP
antennas, and modulation type.  
Color coding and bold font indicate whether or not the given parameters fit
into the number of 
qubits available on current D-Wave machines.
%The reported predictions on resource requirements are 
%in fact very conservative, as D-Wave has announced that its
%next\hyp{}generation machine will support a qualitative jump in %Chimera connectivity,
%due to the new \emph{Pegasus} graph architecture,
%and other QA devices under manufacture (from MIT Lincoln Lab and %Google)
%will also reduce the overhead generated by the embedding.

%In Appendix Figure~\ref{f:parallel}, we plot the relationship between the 
%parallelization factor and number of qubits for fully\hyp{}connected embeddings in the tested chip.

\paragraph{The embedded version of the Ising problem.}
After embedding into Chimera graph we need to recast the Ising problem into an equivalent problem that has the same ground state, but also satisfies the Chimera graph constraints. We also need to introduce a constant penalty term ($J_F$) to quantify the relatively large coupling that constrains 
%each pair of cloned qubits 
all physical qubits belonging to the same logical qubit
to prefer the same state. Appendix~\ref{s:emb-ising} contains additional detail, but we discuss 
important experimental considerations for choosing $J_F$ in Section~\ref{s:eval:thermal}.

%\begin{figure}
%\centering
%\includegraphics[width=1.0\linewidth]{figures/Embedded_HJ.png}
%\caption{\normalfont Absolute Maximum Embedded Ising Coefficients. {\it Left:} diagonal Ising H, {\it Right:} non-diagonal Ising J. }
%\label{f:embedded_max_hj}
%\end{figure}
\parahead{Unembedding with majority voting.}
The bit string that the DW2Q returns is expressed in terms
of the embedded Ising problem, %Eq.~(\ref{eq:IsingEmbeddedObjFunJF}--\ref{eq:IsingEmbeddedObjFun2})
and so must be \emph{unembedded} in order to have 
the values of the bits expressed in terms of our ML
Ising problem.
This is done by checking that all the 
qubits of a logical chain 
are either $+1$ or $-1$.  Should not all spins be concordant,
the value of the 
corresponding logical variable is
obtained by \emph{majority voting} 
(in case of a vote tie, the value is randomized).
Once the logical variables are determined, each configuration yields the corresponding energy of the Ising objective function by substituting it into the original Ising spin glass equation (Eq.~\ref{eqn:ising}).

%The extent to
%which correctly solving a given QUBO problem instance is termed
%\emph{resilience} in the literature.
%Another related way of expressing this problem is to say that the effective
%arithmetic precision that today's quantum annealing platforms have
%available to them is typically rather modest:\ for example, ARC's
%current D-Wave 2000Q annealer has a practical operating 
%arithmetic precision of 
%roughly eight bits (taking values in the range $[-16,16]$ with
%approximately one decimal digit of precision).

%{\fromDV{I commented out in Latex the figure EmbeddedHJ.png 
%that is related to this discussion.
%I don't understand it, not sure what message we want to give.
%If we want to keep the issue, we should comment here on the precision issues but I don't
%understand how is it possible that we have coefficients of
%BPSK so large. Also I don't understand how is this related to
%the best JFerro found, what value of JF is used in the plots?}}

%\subsection{Runs and Performance}
%\label{s:solution_quality}
\section{Implementation}
\label{s:impl}
\phantomsection
\label{s:impl:flow}
\begin{figure*}
    \centering
    \includegraphics[width=0.97\linewidth]{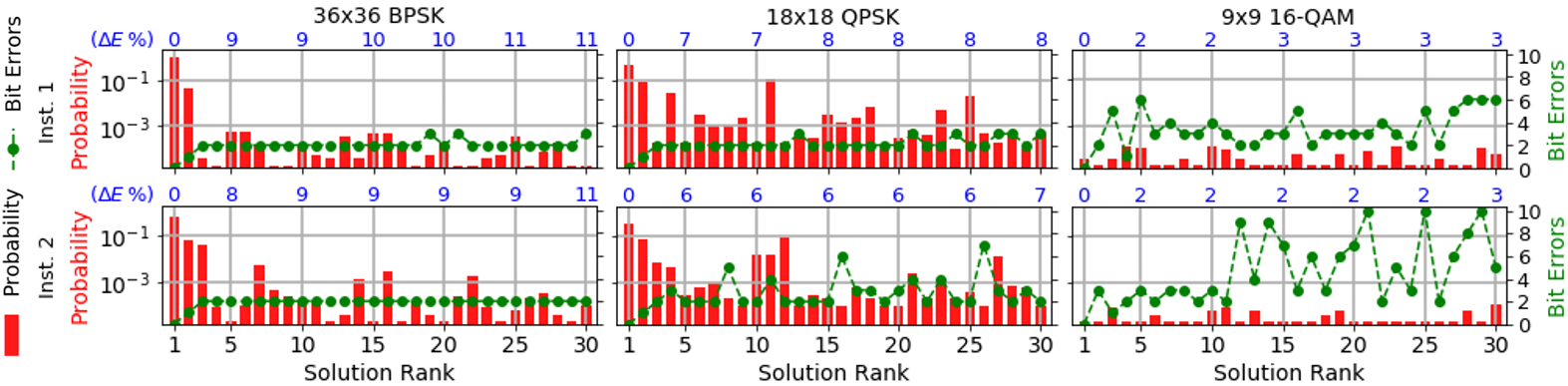}
    \caption{\normalfont Empirical QA results from six different
decoding problems, illustrating relationships
    between Ising energy, solution rank, BER.}
    \label{f:count_ising_eg}
\end{figure*}

This section describes our implementation on the D-Wave
2000Q quantum 
annealer (DW2Q), explaining the API between 
the annealer's control plane and
its quantum substrate,
machine parameters, and 
their tuning to the problem at hand.
% \newpage

Each \emph{anneal cycle} on the DW2Q 
yields a configuration of spins ({\it i.e.,} one decoded bit string). 
The user programs the annealer to run a batch of 
$N_{a}$ annealing cycles (one \emph{QA} \emph{run}) with the same 
parameters to accumulate
statistics, which means that we have a set of $N_{a}$ 
configurations from a DW2Q job submission. The lowest energy 
configuration among $N_{a}$ anneals is the best answer found.
%While asymptotically the parallelization factor is just the ratio of the used physical 
%qubits to the
%number of qubits in the chip, in practical chips also a finite\hyp{}size 
%form factor counts. 

\paragraph{Parallelization.} Multiple instances (identical or not) can be run physically alongside each other, reducing run time by the \emph{parallelization factor}\label{def:parallelization_factor}\footnote{While asymptotically the
parallelization factor is just the ratio of used physical 
qubits after embedding to the
number of qubits in the chip $N_{tot}$, in finite\hyp{}size chips,
chip geometry comes into play.} $P_f \simeq N_{tot}/\left(N\left(\left\lceil N/4 
\right\rceil+1\right)\right)$---a small 16\hyp{}qubit
problem employing just 80 physical qubits ({\it e.g.} 16\hyp{}user BPSK, 8\hyp{}user QPSK, and 4\hyp{}user 16\hyp{}QAM) could in fact be run more than 20 times in parallel 
on the DW2Q.

\paragraph{Precision Issues.}
As analog devices, the desired embedded Ising coefficients
(Eqs. \ref{eq:IsingEmbeddedObjFunJF}-\ref{eq:IsingEmbeddedObjFun2} in Appendix~\ref{s:emb-ising})
do not perfectly match their real energy values once
hardcoded in the real machine, and hence give rise
to \emph{intrinsic control errors} (ICE),
\label{def:ice}
an uncontrollable shift in the actual programmed values of
the objective function. ICE is appropriately modeled as
a noise fluctuating at a time scale of the order of
the anneal time, {\it i.e.}, on each anneal, Ising coefficients
are perturbed: $\bf{f_i} \longrightarrow \bf{f_i}+\langle\delta\bf{f_i}\rangle$, 
$\bf{g_{ij}} \longrightarrow\bf{g_{ij}}+\langle\delta\bf{g_{ij}}\rangle$.
where the noise is 
Gaussian with mean and variance 
measured $\langle\delta\bf{f_i}\rangle$ $\simeq0.008\pm0.02$ and $\langle\delta\bf{g_{ij}}\rangle$ $\simeq-0.015\pm0.025$ in the 
most delicate phase of the annealing run~\cite{ICE}.
The impact of ICE on performance depends
on the problem at hand~\cite{PhysRevA.93.012317,PhysRevA.65.012322},
but it is clear that precision issues will arise if the largest
energy scale squeezes the value 
of the coefficients 
(in Eqs.~\ref{eq:IsingEmbeddedObjFun1}--\ref{eq:IsingEmbeddedObjFun2} 
in Appendix~\ref{s:emb-ising}) to a level where ICE  
is likely to erase significant information of the problem's ground state
configuration.

\parahead{Annealer Parameter Setting.}
\label{s:parameter_opt}
\label{s:impl:param}
%There are parameter settings connected to embedding that tune the performance
%due to improving the quantum dynamics and mitigating the detrimental
%impact of the Intrinsic Control Errors (ICE, \emph{i.e.} the finite precision over which the coefficients
%can appear in the QUBO), that if properly tuned could determine order of magnitude differences
%in probability of success \cite{king2014algorithm}.
% (see Figure~\ref{f:expresults})).
% \begin{figure}
% \centering
% \includegraphics[width=1.0\linewidth]{figures/heatmap.png}
% \caption{Ising Distribution. \normalfont Heatmap distribution (Numbers on it are too small)}
% \label{f:heatmap}
% \end{figure}
As discussed in Section \ref{s:embedding}, the $|J_F|$ that enforces a chain of qubits to return a series of
values which are all in agreement (either all $+1$ or $-1$), and the annealing time $T_a$ are both
key performance parameters that 
determine the net time to find a solution, and hence overall
QA performance. We also introduce 1, 10, and 100~$\mu$s \emph{pause time} $T_p$ in the middle of annealing ($T_a=1$~$\mu$s) with various \emph{pause positions} $s_p$, to see the effect of pausing \cite{DWPauseMarshall} on our problems. 
Setting $|J_F|$ too large would 
wash out the problem information due to ICE,
however $|J_F|$ on average should
increase with the number of logical chains in 
fully\hyp{}connected problems in the absence of 
ICE~\cite{PhysRevX.5.031040}. 
Due to the lack of a first\hyp{}principles 
predictive theory on the correct 
value for
a given instance, we present
in Section~\ref{s:eval:thermal} an
empirical investigation 
of the best embedding parameters, 
employing the microbenchmarking methodologies
generally accepted~\cite{Rieffel2015,o15compiling,PhysRevX.5.031040}.
Below we perform a sensitivity
analysis on $J_F$, $T_a$, $T_p$, and $s_p$ (\S\ref{s:eval:thermal:param})
over the ranges $J_F \in \{1.0\hyp{}10.0 \ (0.5)\} $
, $T_a \in \{1, 10, 100\;\mu s\}$, $T_p \in \{1, 10, 100\;\mu s\}$, and $s_p \in \{0.15\hyp{}0.55 \ (0.02)\}$.
%\footnote{Prior work on other similarly embedded problems~\cite{PhysRevX.5.031040} has shown 
%that the deviation in performance from the optimal parameters of $|J_F|$ 
%with respect to using the median optimal value
%across instances parametrized by their number of qubits does not change
%the value of $\text{TTL}(\mathcal{\hat{P}})$ by more than a small prefactor, in the typical case. Hence we believe
%that our approximation constitutes a reasonable estimate of the performance of \systemname{} in
%an operational scenario.}

\parahead{Improved coupling dynamic range.} The \emph{dynamic range} of coupler strengths is defined as the ratio between the maximum and minimum values that can be set ($g_{ij}$ in Eq.~\ref{eqn:ising}). To strengthen interactions between embedded qubits, the DW2Q is able to double the magnitude of valid negative coupler values, effectively increasing the precision 
of embedded problems and reducing ICE. 
However, this \emph{improved range} option, 
when enabled, breaks the symmetry of the Ising
objective function (substituting the opposite signs for connected coefficients 
and their couplings, into the same problem),
and hence precludes averaging over these symmetrical instances
as the DW2Q does without the improved range option, to mitigate leakage errors~\cite{boixo2014evidence}. It is
thus unclear whether the use of this feature is beneficial in the 
end without experimentation, and so we 
%present results 
benchmark
in Section~\ref{s:eval} 
both with and without improved range.

\section{Evaluation}
\label{s:experiments}
\label{s:eval}

We evaluate \systemname{}
on the DW2Q Quantum Annealer
machine shown in Figure \ref{f:dwave}.  We consider the same number of antennas at the clients and AP ({\it i.e.}, $N_t = N_r$).
Section~\ref{s:eval:understanding} introduces QA results, 
while Section~\ref{s:eval:methodology} explains our 
experimental methodology.  After that
in Section~\ref{s:eval:thermal} we
present results under only the presence of the annealer's 
internal thermal noise (ICE).
Sections~\ref{s:eval:chnoise} and \ref{s:eval:tracech} add
wireless AWGN channel noise and trace\hyp{}based 
real\hyp{}world wireless channels, respectively,
quantifying their interactions
with ICE on end\hyp{}to\hyp{}end
performance.
Over $8 \times 10^{10}$
anneals are used in our performance evaluation.

\subsection{Understanding Empirical QA Results}
\label{s:eval:understanding}

We begin with a close look at two runs of the QA
machine, to clarify the relationships between Ising energy,
Ising energy\hyp{}ranked solution order, and BER.  Figure~\ref{f:count_ising_eg}
shows six QA problem instances (all of which require
36 logical qubits), corresponding
to two different wireless channel uses for each of varying
modulation and number of users.  
The solutions are sorted (ranked) by their \emph{relative Ising
energy difference} $\Delta E$ 
from the minimum  
Ising energy
(blue numbers atop selected solutions), 
where red bars show each solution's frequency of occurrence 
(in the rare case of 
two or more tied distinct solutions, we split those solutions 
into multiple solution ranks).
The number of bit errors associated with each 
solution appears as a green curve. 
For statistical significance,
this data summarizes
50,000 anneals, more than \systemnames{} practical
operation.
As modulation order 
increases and number of users decreases (from left to right in Figure~\ref{f:count_ising_eg}), 
the probability of finding the ground state tends 
to be lower, while the 
search space size remains constant, leading eventually to higher
BER and FER.\footnote{See section \ref{s:eval:methodology_ber}. Frame error rate FER is 
computed as $1-(1-BER)^{\text{frame size}}$.} 
The relative Ising energy gap also
trends smaller,\footnote{The energy 
distribution of the Ising objective function 
(Eq.~\ref{eqn:ising}) corresponds to the distribution of ML decoder Euclidean distances (Eq.~\ref{eqn:ml}).} 
and is likely to be inversely correlated with the
impact of ICE on the problem instance \cite{PhysRevA.93.012317,2018arXiv180603744A}.

\phantomsection
\subsection{Experimental methodology}
\label{s:eval:methodology}

%\begin{figure*}
%\includegraphics[width=1.0\linewidth]{figures/qqq_update.png}
%\caption{The bit error count distribution (averaged over 10 instances) 
%for different modulations with a random channel. 
%As problem size increases, the probability of finding the ground state (zero bit errors) 
%tends to lessen. 
%For the same size of qubits, which implies the same 
%search space size for QA, the results are better 
%in order of BPSK, QPSK, 16-QAM, and finally 64-QAM modulation.}
%\label{f:error_spread}
%\end{figure*}

In this section, we introduce  
performance metrics and figures of merit that give
insight into the operation of the QA machine as it
solves the ML MIMO decoding problem.

We note that in our performance evaluation
we exclude from consideration
programming time and post\hyp{}programming time of the Ising coefficients on the chip, and 
readout latency of the qubit states after a single anneal. 
Currently, these times dominate the pure computation time ({\it i.e.} total anneal time) by several 
orders of magnitude (milliseconds), due to engineering limitations of the technology. 
However, these overheads do not
scale with problem size and 
are not fundamental performance 
factors of the fully integrated \systemname{} system, and so
this is in accordance with experimental QA literature convention. We discuss these overheads in Section~\ref{s:Discussion}.

\subsubsection{Metric: Time-to-Solution (TTS)}
\label{s:eval:methodology_tts}

Suppose we find the ground state (corresponding to 
the minimum energy solution within the search space 
of $2^N$ bit strings, where $N$ is the variable count) 
with probability $\mathcal{P}_0$.  In the absence (but not
presence) of channel
noise, this ground state corresponds to a correct decoding.
Each anneal is an independent, identically\hyp{}distributed
random process, meaning that the expected 
\emph{time to solution}, or $TTS(\mathcal{P})$, 
is the anneal time of each anneal $T_{a}$ multiplied by
the expected number of samples to be able to find the 
ML solution with probability $\mathcal{P}$:
$TTS({\mathcal{P}}) = T_a\cdot 
\log(1-\mathcal{P})/\log(1-\mathcal{P}_0)$.
TTS is commonly used in the QA literature, 
setting $\mathcal{P}=0.99$
\cite{ronnow2014defining}.

%% 1us for both
% \begin{figure}
% \centering
% \includegraphics[width=1.0\linewidth]{figures/annealtime_1us.png}
% \caption{\normalfont Time-to-Solution comparison across different strengths of ferromagnetic coupling within logical qubits, J-ferros ($|J_{F}|$). {\it Left:} BPSK (24-, 36-, 48-, and 60-Users; - results obtained for $T_a$=1$\mu$ s), {\it Right:} QPSK (12-, 18-, 24-, and 30-Users; results obtained for $T_a$=1$\mu$ s).{\fromDV{check annealing time (Minsung) Correct! I can make both 1us (if you want)}}. Thick line reports the median across 10 random instances, and shadings reports
% the best/worse results.{\fromDV{true?? (Minsung) True!}}}
% \label{f:JFERROopt}
% \end{figure}

\subsubsection{Our Metrics: BER and Time-to-BER (TTB)}
\label{s:ber}
\label{s:eval:methodology_ber}

TTS reflects the expected time to find the ground state, 
but does not characterize the 
expected time our system takes to achieve 
a certain \emph{Bit Error Rate} (\emph{BER}, averaged across users),
the figure of merit at the physical layer.
This quantity differs from TTS, because TTS only considers the
ground state, and as illustrated in the example 
shown in Figure~\ref{f:count_ising_eg}, solutions
with energy greater than the ground state may also have (rarely) no or 
relatively few bit errors, while wireless channel
noise may induce bit errors in the ground state solution itself.
Hence we introduce a
metric to characterize the time required 
to obtain a certain BER $p$, \emph{Time-to-BER}: \emph{TTB(p)}.
This is preferred in our 
setting, since a low but non\hyp{}zero bit
error rate is acceptable (error control coding operates 
above MIMO detection).

% \begin{table}[]
%     \centering
%     \caption{Example---Solutions (of each anneal) from two instances, corresponding
%     to two wireless channel uses.  The solutions are ranked by their Ising
%     energy. Ground\hyp{}truth bit error count, and occurrence count of each is shown.}
%     \label{t:energy_dist_eg}
%     \begin{small}
%     \begin{tabular}{l*{6}{r}} \toprule
%     \rowcolor{light-blue}\multicolumn{2}{l}{{\bf Solution rank} ($r$)}& 1& 2& 3& 4\\ \midrule
%     \multirow{3}{*}{\rotatebox[origin=c]{90}{{\bf Inst.~1}}}
%          & {\bf Ising energy}& $-151$& $-145$& $-144$& $-124$ \\
%          & {\bf Bit error count ($F_1$)}& $0$& $1$& $2$& $1$\\
%          & {\bf Occurrences}& $300$& $350$& $100$& $250$ \\ 
%          \midrule
%     \multirow{3}{*}{\rotatebox[origin=c]{90}{{\bf Inst.~2}}}
%          & {\bf Ising energy}& $-120$& $-100$& $-88$&\\
%          & {\bf Bit error count ($F_2$)}& $0$& $3$& $1$\\
%          & {\bf Occurrences}& $600$& $250$& $150$\\ 
% %         \midrule
% %    \multirow{2}{*}{\rotatebox[origin=c]{90}{{\bf Avg.}}}
% %         & {\bf Bit error count ($F_{avg}$)}& $0$& $1.73$& $1.4$& $1.0$\\
% %         & {\bf Occurrences}& $450$& $400$& $125$& $125$\\ 
%     \bottomrule
%     \end{tabular}
%     \end{small}
% \end{table}

\paragraph{TTB for a single channel use.}
Since one QA run 
includes multiple ($N_a$) anneals, we return the 
annealing solution with minimum energy among all anneals 
in that run.  We show an example of this process for one
\emph{instance}
(\emph{i.e.}, channel use, comprised of certain
transmitted bits and a certain wireless channel) in Fig.~\ref{f:count_ising_eg}.
%at the top of Table~\ref{t:energy_dist_eg}.  
The annealer finds different solutions, with different 
Ising energies, ranking them in order of their energy.
Considering this order statistic, 
and the fact that \systemname{} 
considers only the best solution found by all the anneals
in a run, the 
\emph{expected} BER of instance $I$ after $N_a$ anneals
can be expressed as
\begin{small}
\begin{align}
\mathbb{E}(BER(N_a)) = \sum^{L}_{k=1} 
    \left[\left(\sum^{L}_{r=k}p_I(r)\right)^{N_{a}} - 
    \;\left(\sum^{L}_{r=k+1}p_I(r)\right)^{N_{a}}\right]\cdot F_I(k)/N,
\label{eqn:BER}
\end{align}
\end{small}
where $N$ is qubit count, 
$L$ ($\leq N_a$) is the number of distinct solutions,
$r$ ($1 \leq r \leq L$) is the rank
index of each solution, $p(r)$ is the probability of 
obtaining the $r^{\mathrm{th}}$ solution, and $F_I(k)$ is 
the number of bit errors of the $k^{\mathrm{th}}$
solution against ground truth.\footnote{Note that
the metric has omniscient knowledge of ground
truth transmitted bits, while the machine does not.}
To calculate TTB($p$), we replace the left hand side of Eq.~\ref{eqn:BER} with $p$, solve for $N_a$, 
and compute TTB($p$) $=N_aT_a/P_f$.
%($P_f$ is in p.~\pageref{def:parallelization_factor})

\paragraph{Generalizing to multiple channel uses.} The preceding
predicts TTB for a fixed instance.  In the following study we compute
TTB and BER across multiple instances (random transmitted bits 
and randomly\hyp{}selected wireless channel), 
reporting statistics on the resulting sampled distributions.

\subsection{Performance Under Annealer Noise}
\label{s:eval:thermal}

This section presents results from the DW2Q annealer for
wireless channel noise\hyp{}free scenarios, in order to
characterize the machine's performance itself as a 
function of time spent computing.
Sections~\ref{s:eval:chnoise} and~\ref{s:eval:tracech}
experiment with Gaussian noise and trace\hyp{}based
wireless channels, respectively.

% \begin{figure}
% \centering
% \includegraphics[width=0.85\linewidth]{figures/relative_error_cdf.png}
% \caption{CDF of relative Ising energy error for instances with 36 qubits, for different modulations.} 
% \label{f:enrgy_cdf}
% \end{figure}

In this section, we run several
instances with unit fixed channel gain and average 
transmitted power. Each instance has a random\hyp{}phase channel,
randomly chosen transmitted bit string, and is repeated
for each of three different modulations (BPSK, QPSK, 
16-QAM) and varying numbers of users and AP antennas.  
Each instance is reduced to Ising as described in 
Section~\ref{s:reducing},
for a total of 780 different problems
per QA parameter setting. Unless otherwise
specified, this and subsequent sections use the fixed parameter
settings defined in \S\ref{s:eval:thermal:param}.
We obtain significant statistics
by postprocessing up to 50,000 anneals per QA run (except 10,000 anneals for anneal pause analysis in Figure~\ref{f:anneal_pause}).
%and reported figures are based on \emph{Opt($J_F$,$s_p$)} parameter setting in Sec~\ref{s:eval:thermal:param}.

\begin{figure}
\centering
\includegraphics[width=0.90\linewidth]{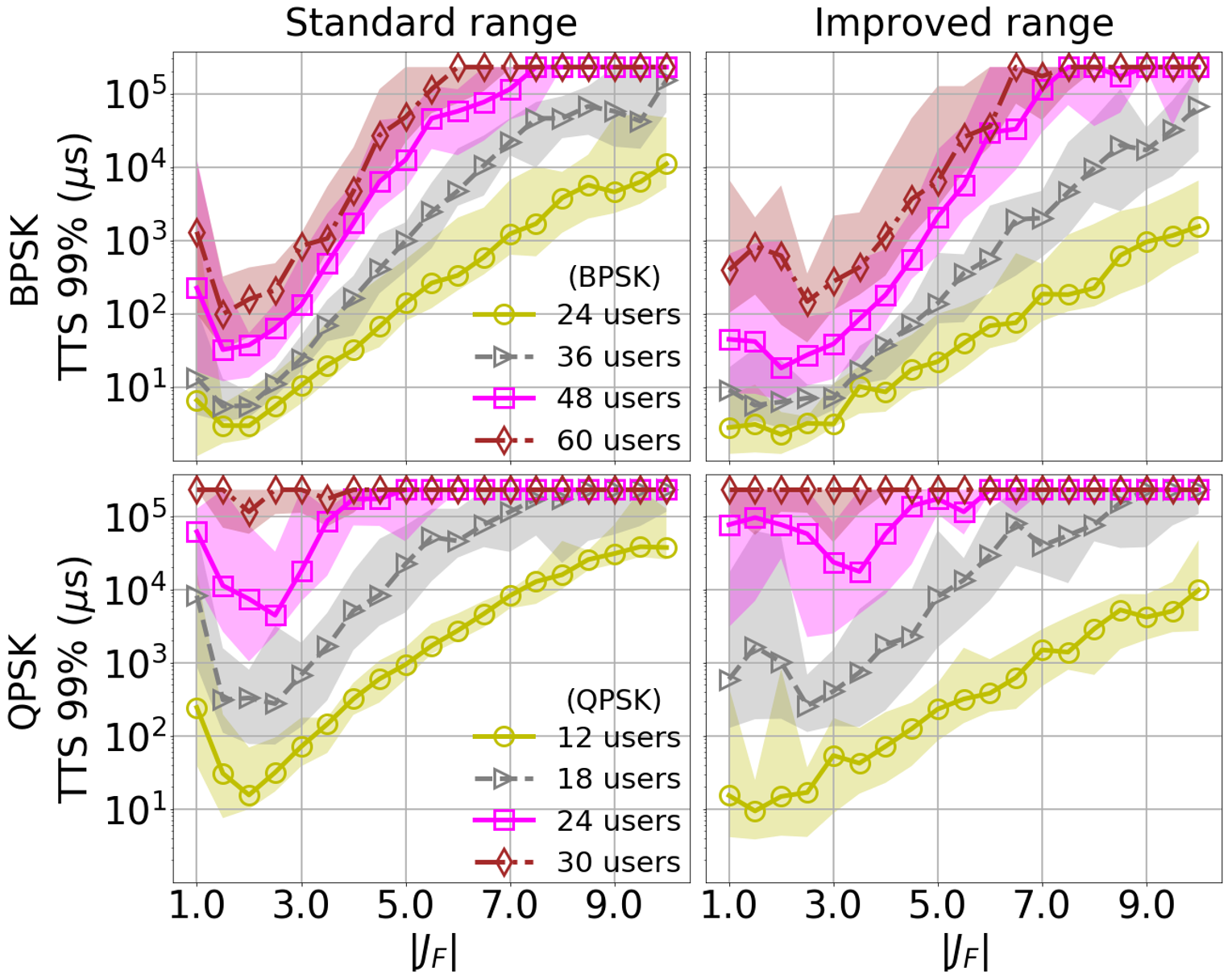}
\caption{\normalfont Time-to-Solution comparison of different strengths of ferromagnetic coupling within logical qubits, $|J_{F}|$. {\it Upper:}~BPSK, {\it lower:} QPSK, {\it left:} standard range, {\it right:} improved range; results obtained for $T_a=1$~$\mu$s. Lines report median of 10 instances; shading reports
10th. and 90th. percentiles.}
\label{f:JFERROopt}
\end{figure}

\begin{figure}
\centering
\includegraphics[width=0.96\linewidth]{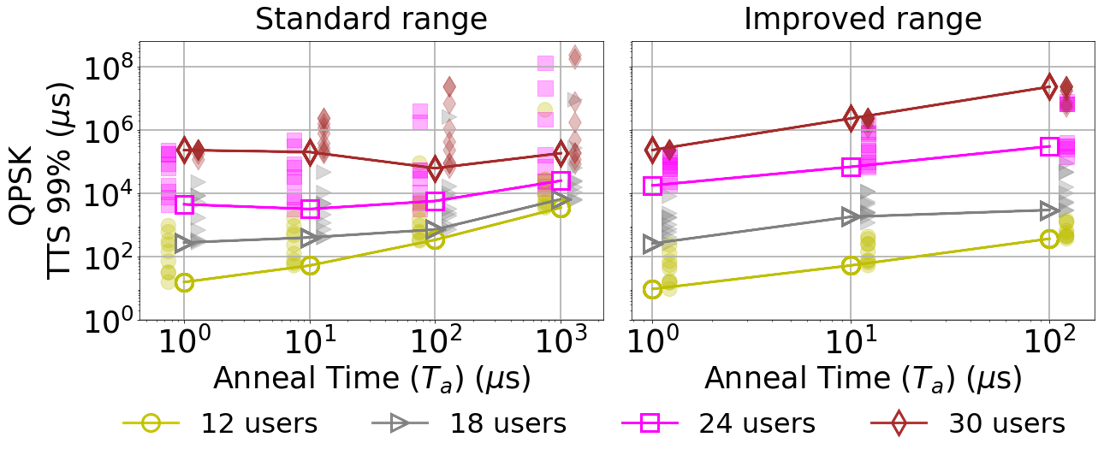}
\caption{\normalfont TTS analysis of different anneal times for different
numbers of users, for QPSK.  Scatter points correlate median results obtained 
for different $|J_F|$, while lines highlight the 
best $|J_F|$ measured from Fig.~\ref{f:JFERROopt} and reporting the median across 10 random instances.}
\label{f:anneal_time}
\end{figure}

\subsubsection{Choosing Annealer Parameters}
\label{s:eval:thermal:param}
In order to isolate the effect of different parameter settings on individual problems, we employ microbenchmarks on TTS. This section explains our choice of parameter settings for our main performance results in \S\ref{s:eval:thermal:ttb}, \S\ref{s:eval:chnoise}, and \S\ref{s:eval:tracech}. 
% We test performance sensitivity for J-ferros $|J_F|$, anneal time $T_a$, pause time $T_p$, pause position $s_p$, and improved coupling dynamic range, as discussed in \S\ref{s:impl}.
Note that while we plot results here only for BPSK and QPSK to save space, our results show that the methods, arguments and observations generalize to higher modulations, unless otherwise indicated. For the purpose of setting the parameters, we restrict the dataset to the ML problems that solve within a median TTS(0.99) of 10~ms for which we have low uncertainty on the measured success probability. We use the determined parameters for all instances regardless of their TTS for the performance analysis.

%In this section we restrict our analysis to instances with statistically relevant time interval (1,000~s) for results, but assume that only the ML problems that reach median TTS(0.99) within 10~ms are valid for further analysis.}

\begin{figure}
\centering
\includegraphics[width=0.96\linewidth]{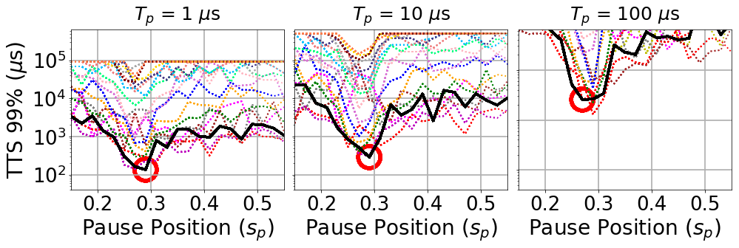}
\caption{\normalfont TTS analysis of anneal pause time and position for 18\hyp{}user QPSK. Colored dotted lines join results obtained 
for different $|J_F|$, while the thicker black line highlights the 
best $|J_F|$ measured from Fig~\ref{f:JFERROopt}. Lines report the median across 10 random instances. The red circle indicates the best $s_p$ for chosen $|J_f|$.}
\label{f:anneal_pause}
\end{figure}

\parahead{Ferromagnetic couplings:}
We examine median 
$\text{TTS}(0.99)$ versus $|J_F|$
over 10 random instances of 
different sizes both with and without extended dynamic range.
In Fig.~{\ref{f:JFERROopt}}, we observe that while there
is a performance optimum that depends on the problem size for standard dynamic range, extended dynamic range shows less sensitivity to $|J_F|$, obtaining roughly the optimal $|J_F|$ performance of standard dynamic range. 
{\bf Anneal time:} As we vary $T_a$, we observe greater sensitivity
when we use non\hyp{}optimal $J_F$, as the scatter points next to each data point in Fig.~\ref{f:anneal_time} (\emph{left}) show.  On the other hand, Fig.~\ref{f:anneal_time} (\emph{right}) shows that
an extended dynamic range setting achieves best results at $T_a=1$~$\mu$s regardless of problem size, showing less sensitivity to different $|J_F|$.  
{\bf Anneal Pause Time and Location:} When we apply
improved dynamic range at $T_a=1$~$\mu$s, we observe a slight independence
(Fig.~\ref{f:anneal_pause}) of $s_p$  and $J_f$ on $T_p$, and as $T_p$ increases, so does TTS. 
While the dynamic range setting has shown less sensitivity to $|J_F|$, anneal pause with extended dynamic range shows more sensitivity. Note that TTS of 18\hyp{}user QPSK at $T_p=1~\mu$s is slightly improved, compared to the best results in Figs.~\ref{f:JFERROopt}~and~\ref{f:anneal_time}.

%Note that TTS of 18\hyp{}user QPSK at $T_p=1~\mu$s is slightly improved, compared to the best results in Figs.~\ref{f:JFERROopt} and ~\ref{f:anneal_time}.

\paragraph{Annealer Parameter Optimization.}
Based on the previous sensitivity analysis, we select a default QA parameter setting. First, we choose {\bf improved dynamic range} since it is relatively robust to choice of $|J_F|$, nearly equaling the best performance of the standard dynamic range. Second, we choose $\mathbf{ T_p=1~}$ $\bm{\mu}$\textbf{s}, since it shows better results and greater pause times dominate the anneal time.

\subsubsection{Choosing whether to pause}
\label{s:eval:whetherpause}
With the above default QA parameters, we now use TTB to explore whether or not we should use the QA pause functionality, as TTB encompasses both algorithms' BER performance 
%\sout{(which TTS fails to do)} 
as well as wall clock running time (\emph{cf.} TTS). 
%Note that these two different optimization is to see non-pause scenario with optimized $J_F$, $T_a$ \emph{versus} pause scenario with optimized $J_F$, $T_a$, which is a meaningful comparison in that the anneal pause might bring the better probability of finding ground state but it could take more time. 
We first define a \emph{fixed parameter setting} by selecting the best estimated choices for the non\hyp{}pausing algorithm and for the pausing algorithm, meaning the parameters which optimize medians across a sample of instances belonging to the same problem class ({\it e.g.} 18$\times$18 QPSK).
%, denoted as \textbf{\textit{Fix($\bm{J_F}$, $\bm{T_a}$)}} and \textbf{\textit{Fix($\bm{J_F}$, $\bm{s_p}$)}} respectively, in order to evaluate \systemnames{} practical performance. 
%We also compare 
This approach is to be compared 
against an \emph{oracle} that optimizes \{$J_F$, $T_a$\} or \{$J_F$, $s_p$\} instance by instance. 
In the figures, we denote the two parameter setting methods as \textbf{\textit{Fix}} (fixed) and \textbf{\textit{Opt}} (optimal),
respectively. 
% Note that our choice of fixed parameter setting for each problem class is done through Fig~\ref{f:JFERROopt}, Fig~\ref{f:anneal_time}, and Fig~\ref{f:anneal_pause} ({\it e.g.} 18\hyp{}user QPSK instances are optimized with $|J_F|\simeq 1.5$ (less clear), $T_a \simeq 1\;\mu s$, and $s_p \simeq 0.35\;\mu s$).
\begin{figure}
\centering
\includegraphics[width=0.96\linewidth]{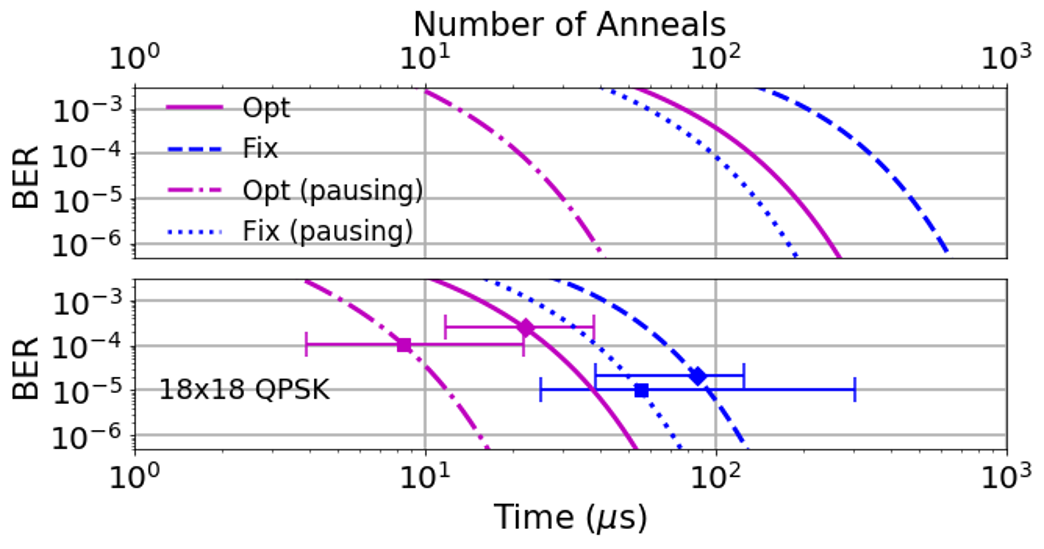}
\caption{\normalfont BER of different optimization settings as a function of the number of anneals ({\it upper}) and time ({\it lower}) for $18\times18$ QPSK (median across 20 instances). Error bars indicate 15th. and 85th. percentiles.}
\label{f:anneal_to_time}
\end{figure}

\begin{figure}
\centering
\includegraphics[width=\linewidth]{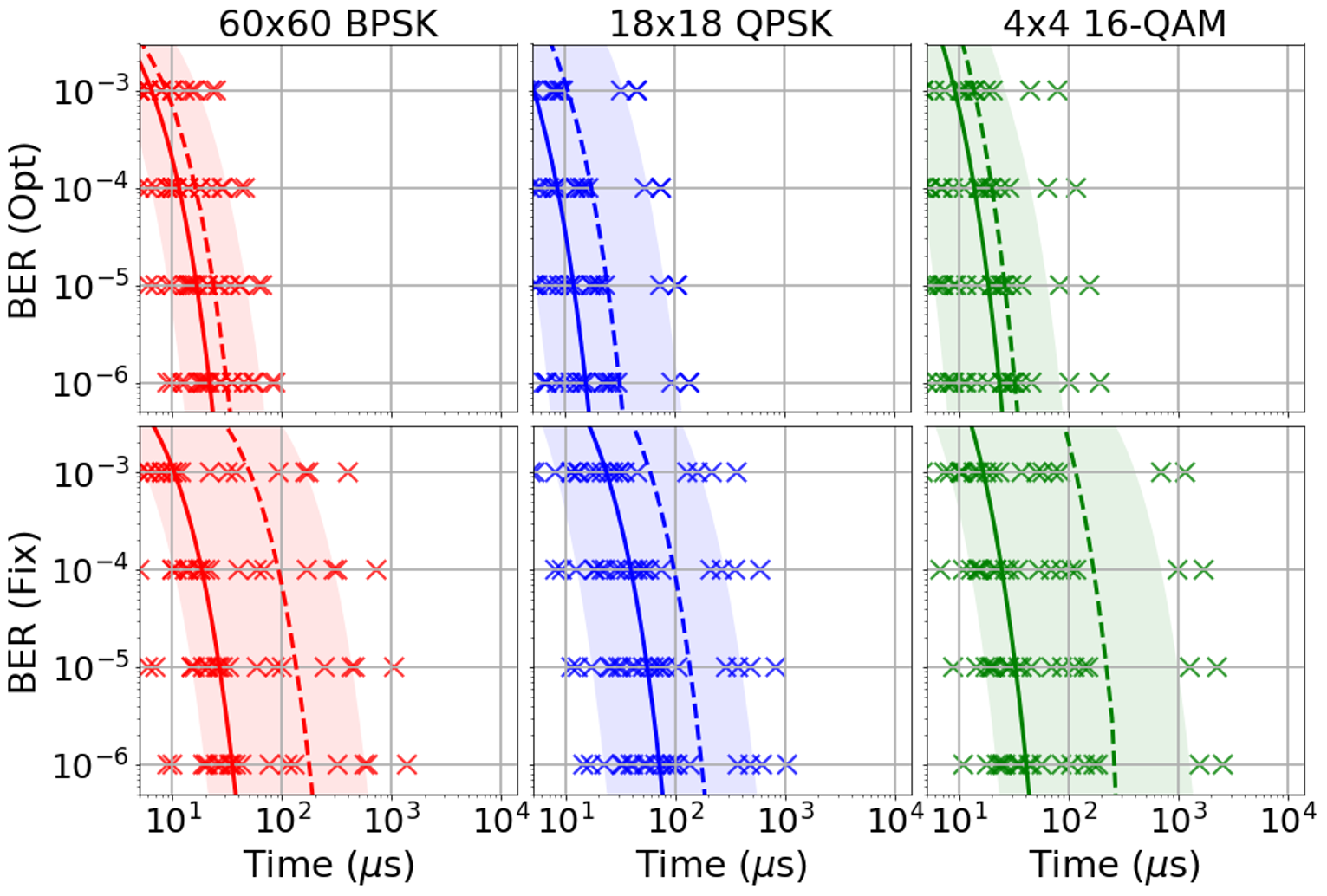}
\caption{\normalfont Time-to-BER (TTB) comparison across different user numbers and modulations. {\it Upper}: ideal scheme using \emph{Opt}.
%optimizing \{$J_F$, $s_p$\} per instance. 
{\it Lower}: \systemnames{} performance optimizing 
with \emph{Fix}.
%\{$J_F$, $s_p$\} per problem class. 
Solid lines and dashed lines report median and mean TTB across 20 instances, respectively. Shading reports 10th. and 90th. percentiles of average BER at a certain time and each $\times$ symbol reports each instance's TTB (x\hyp{}value) for a certain target BER.}
\label{f:ttb}
\end{figure}

Our motivation for considering \emph{Opt} is that it provides a bound to what can be achieved by the methods that seek to optimize machine parameter settings instance by instance~\cite{ReverseVenturelli, venturelli2015quantum}, currently under investigation. With our traces we compute BER as a function of $N_a$ using Eq.~\ref{eqn:BER}; the median result across 20 random instances is shown in Figure~\ref{f:anneal_to_time} ({\it upper}). Figure~\ref{f:anneal_to_time} ({\it lower}) shows the corresponding BER as a function of time ({\it i.e.,} TTB). Note that the pausing algorithm 
%(\emph{Opt} and \emph{Fix}) 
has a better performance than the non\hyp{}pausing algorithm (regardless of \emph{Opt} and \emph{Fix} strategy) despite the fact that each anneal in the former ($T_a +T_p$) takes twice as much time than the latter (when $T_a=1~\mu s$). Based on this empirical finding, we will present the results in the following section only for the protocol that includes a pause.
% This feature has been only observed in unembedded instance sets tested in the literature reporting experiments on D-Wave 2000Q~\cite{albash2017evidence}.  For instance, for the example shown (QPSK modulation) it is clear that the 24 user instances are
% optimized for \emph{Fix($J_F$, $T_a$)} with $|J_F|\simeq 2.5$ and $T_a \simeq 10\;\mu s$.

% We explore two different 2D parameter settings: first experimenting over different $J_F$ and $T_a$ without anneal pause, second experimenting over different $J_F$ and $s_p$ with different pause $T_p$ and $T_a=1 \mu s$, noting that fixed settings provide near\hyp{}optimal
% performance. Some of our experimental results assume an Oracle that optimizes \{$J_F$, $T_a$\} or \{$J_F$, $s_p$\} instance by instance, denoted as \emph{Opt($J_F$, $T_a$), Opt($J_F$, $s_p$)} respectively, but we also compare against
% a fixed (median optimal) \{$J_F$, $T_a$\} and \{$J_F$, $s_p$\} for each problem class ({\it e.g.} 8$\times$8 QPSK), denoted as \emph{Fix($J_F$, $T_a$), Fix($J_F$, $s_p$)} respectively, and note that 
% solution within a factor of 2--5$\times$ of reported
% times to solution is possible by multiple runs and
% $J_F$ settings.

%\newpage

\subsubsection{\systemname{}: End\hyp{}to\hyp{}End performance} 
\label{s:eval:thermal:ttb}

%Figure~\ref{f:error_spread}
%shows the bit error distribution across
%our random instance set, aggregating 
%the distribution of bit errors and reporting average statistics. 
%Results are obtained by annealing at 50,000 times for each instance
%in order to have meaningful statistics. 
%In the case of $24\times 24$ BPSK (Figure~\ref{f:error_spread}, 
%{\it left}), the probability of finding the ground state 
%(corresponding to no bit errors) is about 80\%. 
%Compared to BPSK, different modulations with the same size search space ($=2^{24}$) have a lower probability of finding the ground truth. 
%This indicates that 
%dense Euclidean distance distribution causes 
%dense energy distribution implying many similar 
%local minimums and that precise annealing with 
%high precision is required to distinguish similar magnitudes 
%of energies, to find the lowest energy. As mentioned in Section~\ref{s:embedding},
%the DW2Q has a precision limit, explaining
%the increased number of bit errors at higher 
%order modulations.

% \begin{figure}
% \includegraphics[width=1.0\linewidth]{figures/draft_cdf.png}
% \caption{CDF across instances. \normalfont \textcolor{red}{left (x-axis = BER at 100 anneal), right (x-axis = number of anneals to achieve BER $10^{-6}$) (Rough figure : red 60x60 bpsk, blue 18x18 qpsk, green 4x4 16qam).}}
% \label{f:cdf}
% \end{figure}

\begin{figure}
\centering
\includegraphics[width=\linewidth]{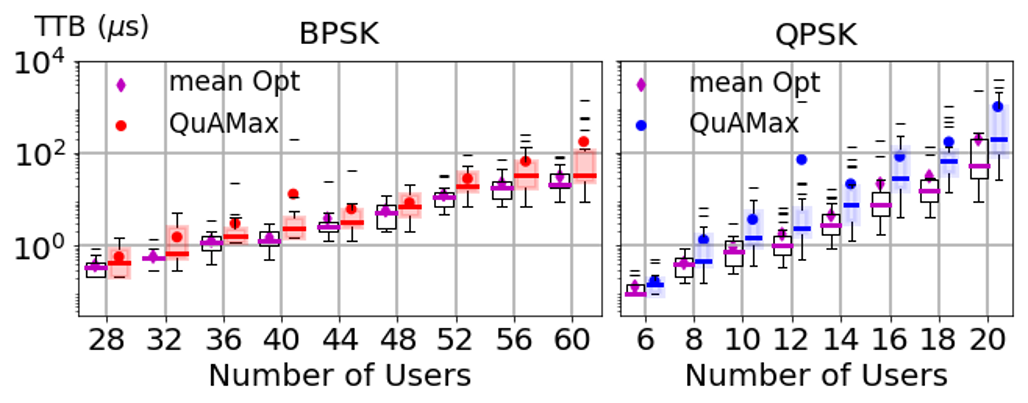}
\caption{\normalfont TTB with target BER $10^{-6}$ for different modulations and user numbers across 20 instances. Colored boxes report \systemname{} 
(5th., 95th.\ as whiskers, upper\fshyp{}lower quartiles
as boxes, median as the thick horizontal mark, and thin horizontal marks for outliers.)}
\label{f:ttbperuser}
\end{figure}

\begin{figure}
\centering
\includegraphics[width=\linewidth]{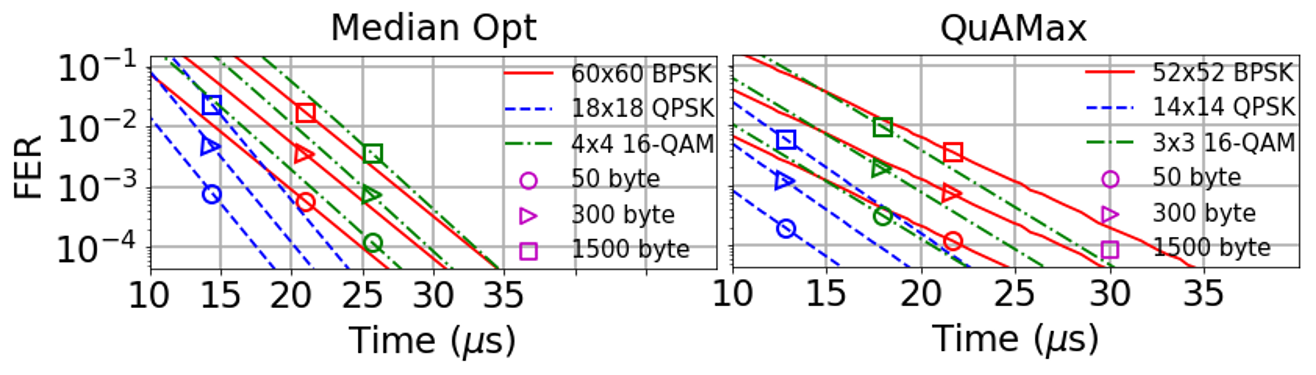}
\caption{\normalfont Time-to-FER for different 
users, modulations, and frame sizes; 
\emph{left}: median Opt (idealized),
\emph{right}: mean \emph{Fix} (\systemname{}).
}
\label{f:fer_baseline}
\end{figure}

We now evaluate the TTB and TTF (Time\hyp{}to\hyp{}FER) of \systemname{}, comparing:
\begin{enumerate}
    \item {\bf \systemname{}:} Fixed\hyp{}parameter, average\hyp{}case performance.
    \item {\bf Oracle:} Median\hyp{}case \emph{Opt} performance
    (\S\ref{s:eval:whetherpause}: outlier data points have minimal influence on
    the median order statistic), optimizing QA 
    parameters.\footnote{Outlier mitigation methods for QA may address
    such outliers in future work~\cite{qubo-preproc}.}
\end{enumerate}
Figure~\ref{f:ttb} shows the TTB with varying user numbers and modulations at the edge of \systemnames{} performance capabilities. Solid and dashed lines report median and average BER, respectively. 
%\textcolor{red}{Both metrics will need to be under control for  \systemname{} to be reliable, but more statistics would be necessary to accurately estimate the mean in the \emph{Fix} case, due to the low precision . For now, we present initial estimates delegating more thorough benchmarks to the future.} 
%Notice that the mean (but not the median) curve of $60\times 60$ BPSK reaches an error floor at just below a BER of $10^{-3}$ indicating that a small but significant number of problem instances have a low probability of finding the ground state.
We note that mean TTB dominates median TTB due to a small number of long\hyp{}running outliers.
\systemname{} accordingly
sets a time deadline (measured median TTB for the target BER) for decoding and after that discards bits, relying on forward error correction to drive BER down.
Next considering the relationship between TTB and problem size, Fig.~\ref{f:ttbperuser} explores TTB for target BER $10^{-6}$, for each instance that reaches a BER of $10^{-6}$ within 10~ms   
as well as average performance. 
%We observe that as the number of users increases, TTB increases with a relationship that is largely independent of the modulation chosen, in contrast to conventional ML decoders.
ML problems of these sizes are well beyond the capability of
conventional decoders (\emph{cf.} Table~\ref{t:visited_nodes}), and we observe that \emph{Opt} achieves superior BER within 1--100~$\mu s$ and that \systemname{} achieves an acceptable BER for use below error control coding. %\textcolor{red}{in a future version of the method we could introduce} preprocessing on the Ising form to improve performance and reduce outliers~\cite{qubo-preproc}. 
Note that instances with TTB below the minimum required time ({\it i.e.,} $T_a$ + $T_p$) caused by parallelization require (an amortized) 2~$\mu s$.

Next, we consider frame error rate performance, 
measuring mean and median FER \systemname{} achieves.
Results in Fig.~\ref{f:fer_baseline} show that tens 
of microseconds suffice to achieve a low enough (below $10^{-3}$)
FER to support high throughput communication 
for 60\hyp{}user BPSK, 18\hyp{}user QPSK, or 
four\hyp{}user 16\hyp{}QAM suffices to serve four 
users with the idealized median performance of \emph{Opt}. \systemname{} (mean \emph{Fix}) achieves a similar performance 
with slightly smaller numbers of users.
Furthermore, our results show low sensitivity to frame 
size, considering maximal\hyp{}sized internet data
frames (1,500~bytes) all the way down to TCP ACK\hyp{}sized 
data frames (50~bytes).

\subsection{Performance under AWGN Noise} 
\label{s:eval:chnoise}

\begin{figure}
\centering
\includegraphics[width=0.99\linewidth]{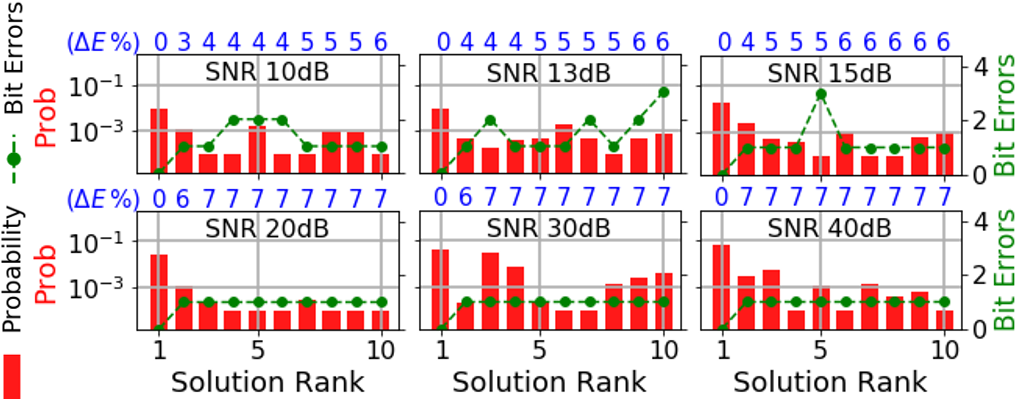}
\caption{\normalfont Detailed view (\emph{cf.} Fig.~\ref{f:count_ising_eg}) of an example wireless channel at six different SNRs  (18\hyp{}user QPSK).}
\label{f:SNR_anlaysis}
\end{figure}

\begin{figure}
\centering
\includegraphics[width=1.0\linewidth]{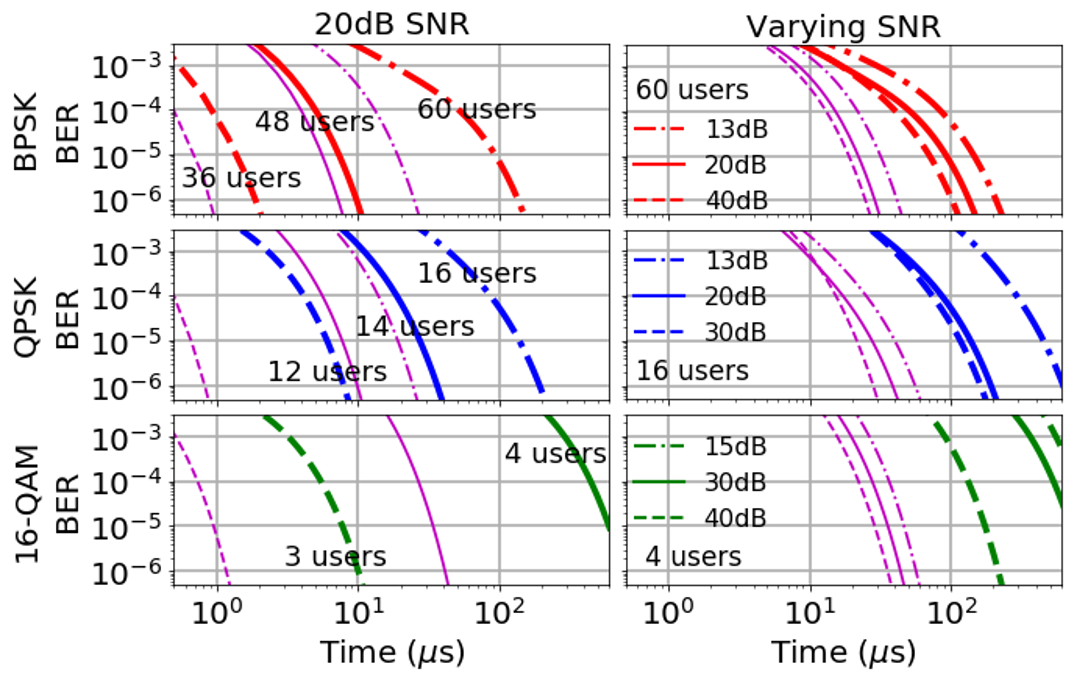}
\caption{\normalfont TTB comparison across different user numbers, modulations, and SNRs. {\it Left}: varying the number of users at SNR 20~dB. {\it Right}:~varying SNR at a certain number of users. Thick lines report \systemnames{} performance (mean \emph{Fix}), and same style but thin (purple) lines report the idealized performance (median \emph{Opt}).} 
\label{f:diffsnr}
\end{figure}

We next evaluate the impact of AWGN from 
the wireless channel,
testing six SNRs ranging from 10~dB to 40~dB. In order to isolate the effect of noise, the results in this subsection fix the channel and transmitted bit\hyp{}string and consider ten AWGN noise instances. Looking at the data in depth to begin with, the effect of AWGN channel noise, which is itself additive to ICE,
is shown in Fig.~\ref{f:SNR_anlaysis} for six illustrative examples. As SNR increases, the probability of finding the ground state and the relative energy gap tend to increase. At 10~dB SNR the energy gap between the lowest and second lowest energy solutions narrows to just three percent, leaving minimal room for error. In terms of overall performance, Fig.~\ref{f:diffsnr}~(\emph{left}) shows TTB at 20~dB SNR, varying number of users and modulation. At a fixed SNR, we observe a graceful degradation in TTB as the number of users increases, across all modulations. Fig.~\ref{f:diffsnr}~(\emph{right}) shows TTB at a certain user number, varying SNR and modulation. At a fixed user number, as SNR increases, performance improves, noting that the idealized median performance of \emph{Opt} shows little sensitivity to SNR, achieving $10^{-6}$ BER within $100~\mu$s in all cases.

\begin{figure}
\centering
\includegraphics[width=1.0\linewidth]{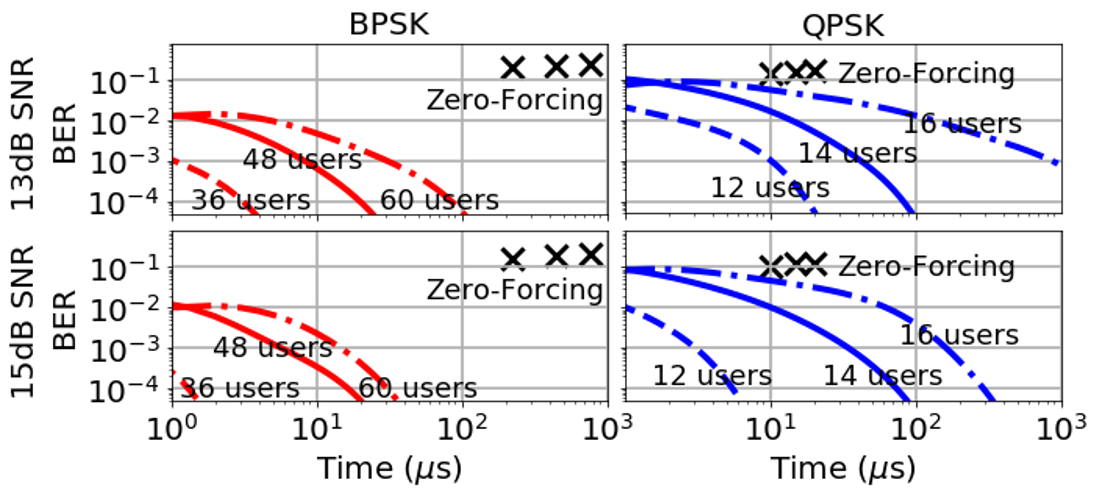}
\caption{\normalfont \systemnames{} performance comparison against the zero-forcing decoder across different user numbers, modulations, and SNRs. Each $\times$ symbol (\emph{left\hyp{}to\hyp{}right:} 36, 48, 60 users for BPSK and 12, 14, 16 users for QPSK) reports the zero-forcing decoder's BER and corresponding processing time.}
\label{f:comparison}
\end{figure}

Fig.~\ref{f:comparison} compares \systemnames{} performance versus zero-forcing decoder at bad SNR scenarios, showing the necessity of ML-based MIMO decoders for large MIMO system. Linear decoders such as zero-forcing and MMSE suffer from the effect of the poor channel condition (when $N_t\approx N_r$), requiring $N_r > N_t$ ({\it i.e.,} more antennas) for appropriate BER performance. In Figure~\ref{f:comparison}, \systemname{} reaches the zero-forcing's BER (or even better BER) approximately 10-1000 times faster than zero-forcing in both BPSK and QPSK modulation. Here, computation times for zero-forcing are inferred from processing time using a single core in BigStation \cite{BigStation}, one of the current large MIMO designs based on zero-forcing. While this processing time can be reduced proportionally with more cores, BER ({\it i.e.,} quality of solutions) remains unchanged. The Sphere Decoder achieves comparable BER, but processing time cannot fall below a few hundreds of $\mu$s with the given numbers of users and SNRs.\footnote{Extreme levels of parallelization or GPU implementation might be able to resolve the issue. However, practical constraints 
%\textcolor{red}{(even on pure computation \cite{hill2008amdahl})} 
will 
%continue to limit speedup 
eventually limit the increase in performance on classical platforms \cite{king2019quantum}. Contrarily, overheads in \systemname{} are apart from pure computation, which can be resolved by engineering design.}
%gains 
%on classical platforms \cite{king2019quantum}. 

\subsection{Trace-Driven Channel Performance}
\label{s:eval:tracech}

\begin{figure}
\centering
\includegraphics[width=0.95\linewidth]{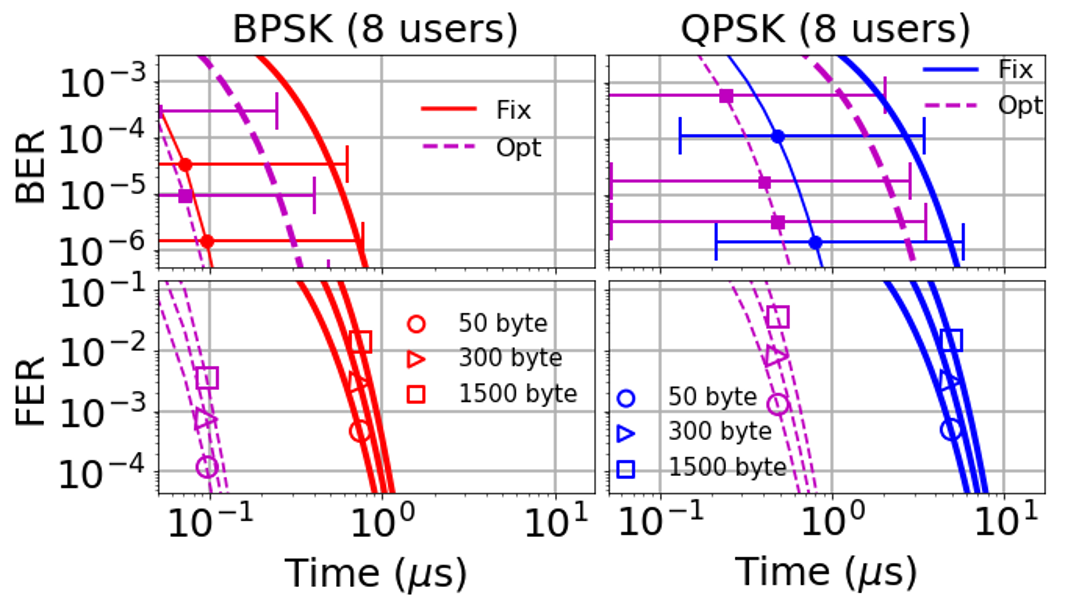}
\caption{\normalfont Experimentally measured channel trace~\cite{Argos} results: upper plots report TTB (\emph{Opt}, \emph{Fix}); lower plots report TTF of median \emph{Opt} and mean \emph{Fix} (QuAMax). Thin and thick lines report median and mean, respectively. TTB's error bars indicate 15th.\ and 85th.\ percentiles.}   
\label{f:actual_trace}
\end{figure}

We evaluate system performance
with real wideband MIMO channel traces at 
2.4~GHz, between 96 base station antennas and eight static users \cite{Argos}.
This dataset comprises the largest MIMO trace size currently available. For each channel use, we randomly pick eight base station antennas to evaluate the $8 \times 8$ MIMO channel use at SNR {\it ca.} 25--35~dB: Fig.~\ref{f:actual_trace} shows the resulting TTB and TTF. We achieve $10^{-6}$ BER and $10^{-4}$ FER within $10~\mu$s for QPSK. For BPSK, considering multiple problem instances operating in parallel, we achieve the same BER and FER within (an amortized) $2~\mu$s period.
This implies that tens of microseconds suffice to achieve a low BER and FER even without parallelization of identical problems, which creates an opportunity for \systemname{} to parallelize different problems (\emph{e.g.}, different subcarriers' ML decoding).

% \begin{figure*}
%     \centering
%     \begin{subfigure}[b]{0.3\linewidth}
%     \centering
%     \includegraphics[width=\linewidth]{figures/24.png}
%     \caption{24 qubits.}
%     \label{f:qubo_spread:8by8}
%     \end{subfigure}
%     \hfill
%     \begin{subfigure}[b]{0.3\linewidth}
%     \centering
%     \includegraphics[width=\linewidth]{figures/36.png}
%     \caption{36 qubits.}
%     \label{f:qubo_spread:8by8logy}
%     \end{subfigure}
%     % \begin{subfigure}[b]{0.4\linewidth}
%     % \centering
%     % \includegraphics[width=\linewidth]{figures/48.png}
%     % \caption{48 qubits.}
%     % \label{f:qubo_spread:8by8logy}
%     % \end{subfigure}
%     \hfill
%     \begin{subfigure}[b]{0.3\linewidth}
%     \centering
%     \includegraphics[width=\linewidth]{figures/60.png}
%     \caption{60 qubits.}
%     \label{f:qubo_spread:8by8logy(1-CDF)}
%     \end{subfigure}
% \caption{The spread of error bits for different modulations. As problem (qubit) size increases, probability of finding ground truth, where error bit is 0, tends to be lower.}
% \label{f:qubo_spread}
% \end{figure*}

\section{Related Work}
\label{s:related}

\paragraph{Applications of QA.} 
Despite the immaturity of software toolchains,
existing quantum annealing
machines have been already programmed successfully to 
solve problems in Planning and Scheduling \cite{Rieffel2015}, 
Databases \cite{Trummer:2016:MQO:2947618.2947621},
Fault Diagnostics \cite{Perdomo-Ortiz2015}, 
Machine Learning \cite{perdomo2017opportunities}, 
Finance \cite{Rosenberg:2015:SOT:2830556.2830563}, Data Analysis \cite{mott2017solving}, 
Chemistry~\cite{hernandez2017enhancing}, and
Space Sciences\fshyp{}Aeronautics \cite{biswas2017nasa}. A Similar problem to ML detection, CDMA multiuser demodulation, was solved using quantum fluctuations controlled by the transverse field (similar as QA) in \cite{otsubo2014code}.
Of particular relevance is work on 
optimization of fully\hyp{}connected graphs, such as the ones used to
map the ML problem \cite{PhysRevX.5.031040}; the results of 
which showed that QA performance could match
the most highly optimized simulated 
annealing code run on the latest Intel processors. 
For further details on the logical to physical
qubit embedding process, see Venturelli \emph{et al.}
\cite{PhysRevX.5.031040}.  Efficient embeddings which do 
not force the chip coverage to be a triangle are also 
known~\cite{boothby2016fast}.

%\paragraph{Locality reduction transformations.}
%The appearance of cubic and quartic terms 
%in the optimization function (as described 
%in \S\ref{s:design:higher} is 
%a common occurrence in many combinatorial mappings: 
%in the QA literature, these mappings are dubbed  
%\emph{Polynomial Unconstrained Binary Optimization} (PUBO).
% PUBO to QUBO reductions are available that can be tailored 
%to the problem at hand, \emph{e.g.} \cite{perdomo2017readiness} 
%for PUBO\hyp{}to\hyp{}QUBO mapping of circuit fault detection problems.

\paragraph{QAOA.} Quantum Approximate Optimization Algorithms,
invented in 2014 \cite{farhi2014quantum}, 
and recently generalized for constrained combinatorial optimization \cite{Hadfield:2017:QAO:3149526.3149530},
require digital gate\hyp{}model QC, which became available at reasonable scale 
only in 2017 (prototypes from  IBM, Rigetti Computing, and  Google are available).
While QA and QAOA require
different hardware (the former is analog, the latter 
digital) they have in common that:
    {\bf (1)}~For problems that don't have hard constraints, the programming
    step consists in defining a classical combinatorial problem 
    which is cast into QUBO \cite{boros2007local, QUBO}
    or \emph{Ising} form, hence they both may leverage our 
    formulation \S\ref{s:reducing}.
    {\bf (2)}~QAOA can be seen in some parameter range as a ``digitized'' 
    version of QA, and it has been formally demonstrated 
    that it can simulate the results and performance of QA 
    and outperform it, in principle \cite{yang2017optimizing}.
The first commonality is particularly important 
since it opens the door to application of our techniques
on future hardware capable of running QAOA.

\paragraph{Conventional ML Detectors.}
Faster silicon based ML detector strategies
typically approximate and parallelize
the ML computation \cite{Wenk10, cui-tvt13}. In these general directions, much progress
has been made to the point that
%.  First, approximation
%techniques replace exact calculations with
%computationally simpler but less precise distance measures, to
%incrementally lower processing overhead~\cite{ETH_HARD,
%Hess07,Wenk10}.  A second body of work focuses on the computation
%involved itself, targeting the geometry 
%of the wireless signal constellation and using knowledge 
%of that geometry to reduce the computation burden 
%\cite{Geosphere,Mennenga09,Shabany08}, or probabilistically 
%pruning the search space 
%of the Sphere Decoder \cite{cui-tvt13, cui-icc07, stojnic-icassp06}
%in order to speed
%the overall computation.  A third line of work
%focuses on parallelizing the problem more efficiently, both
%at the level of individual Euclidean distance calculations 
%of received signal constellation points 
%\cite{Wenk10,Mondal10,Shabany08b} and at higher levels
%of the computation, for example running multiple 
%lower\hyp{}dimensional copies of an entire Sphere Decoder itself
%\cite{KhairyGPU}.  
Sphere Decoders have been realized in ASIC
hardware~\cite{ETH_HARD, winter12} but
fall short for the 
reasons noted in Table~\ref{t:visited_nodes} when
the setting demands more antennas at the AP (serving more users), 
or when the modulation chosen increases 
\cite{flexcore-nsdi17, BigStation}.

%\textcolor{red}{\cite{shim-toc10} 8 by 8 16-QAM visits 5,000 nodes (20 us at a 250 MHz clock cycle in 
%an ideal case of one clock cycle per node)}

%{\fromDV{I think here we need more recent references on classical side.
%I see the references cited are few years old.. are they really the state of art?}}

\section{Discussion}
\label{s:Discussion}
In this section, we discuss the current status of QA technology and practical considerations.

\paragraph{Computational Power Consumption.} The computation in the DWQ2 is performed at zero energy consumption, as dictated by reversible computing, although energy is dissipated in the initialization and readout.
%Note that while Sunway
%TaihuLight, one of the most
%powerful supercomputers as of 2019, requires 42~MW of power, 
The DW2Q draws 16~kW of power, primarily used by the cryogenic refrigeration unit~\cite{DWaveWP}. The computational power (per watt) for QPUs is expected to increase much more rapidly than for conventional computing platforms since the DW2Q power draw is not expected to change much as qubit and coupler counts grow in future generation systems while the computational power substantially increases.

\paragraph{Operating Expenses.} Operating the DW2Q results in
significant electricity cost, and the dilution refrigerator requires liquid nitrogen 1-2 times a month, for a total yearly cost of about USD~\$17,000.

\paragraph{Processing Times.} 
%Currently, a latency caused due to the deployment of the QPU as a cloud-based quantum instruction server dominates end\hyp{}to\hyp{}end processing time. 
The scenario envisioned by \systemname{} assumes a centralized RAN architecture where a QPU, co\hyp{}located with centralized RAN computational resources in a data center, is connected to the APs via high\hyp{}speed fiber or millimeter\hyp{}wave links. In this setting, a latency between the APs and data center will not be significant. 
Nonetheless, \systemname{} cannot be deployed today, since additional processing times in the current QPU include approximately 30-50~ms preprocessing time, 6-8~ms programming time, and 0.125 ms readout time per anneal. These overheads are well beyond the processing time available for wireless technologies (at most 3--10~ms). However, these overhead times are not of a fundamental nature and can be reduced by several orders of magnitude by efforts in system integration. By means of extrapolation of improvement trends it is expected that quantum engineering advances in superconducting qubit technology will enable \systemname{} to be viable within a decade.
Moreover, \systemnames{} Ising form (in Section~\ref{s:mltoising}) can be adapted to be run in other emerging physics-based optimization devices based on photonic technologies \cite{hamerly2019} 
whose processing times overhead are in principle much faster.
Hence, we leave an end\hyp{}to\hyp{}end evaluation in a fully centralized RAN architecture, with more advanced hardware, as future work.

% The new system continues D-Wave's record of doubling the number of qubits on its QPUs every two years, which enables larger problems to be run or more parallelization.

\section{Conclusion}
\label{s:concl}

\systemname{} is the
first design, implementation and experimental evaluation
of a quantum\hyp{}computing solver for
the computationally challenging ML MIMO decoding problem.  Our
performance results establish a 
baseline for a future fully-integrated systems
in the context of the centralized RAN architecture.
We show that once engineering efforts 
optimize the integration between quantum and 
conventional computation, quantum computation
should be considered
a competitive technology for the 
future design of high\hyp{}capacity 
wireless networks.

%However, it is clear that even with technological advances on superconducting\hyp{}qubit circuit
%integration, for the whole end-to-end decoding procedure to be performed under 10$\mu s$,
%only a handful of annealing cycles will be affordable, if not just a single one~\cite{abbott2018hybrid}.
%Nevertheless, as discussed in Section \ref{s:embedding}, the anneal procedure is trivially
%parallelizeable so a given required number of repetitions should be cast into a requirement in number of 
%qubits in the chip, by considering that each additional repetition is just additional qubits
%that are being annealed in parallel.

\paragraph{Future Work.} 
There are several improvements over the design we
have evaluated here.  First, we anticipate that 
further optimization of $|J_{F}|$,
$T_a$, and $s_p$ as well as new QA techniques such as 
\emph{reverse annealing} \cite{ReverseVenturelli} 
may close the gap to \emph{Opt} performance.
Second, there are changes in QA
architecture expected in 
annealers due this year
\cite{2019arXiv190107636D} featuring 
qubits with $2\times$ the degree of Chimera, $2\times$
the number of qubits and with
longer range couplings. Based on similar gains 
in recent
results on different problem domains \cite{hamerly2019}, 
we anticipate this will permit ML problems 
of size, \emph{e.g.}
$175\times175$ for QPSK and dramatically increase the parallelization opportunity of the chip
due to the reduced embedding overhead where each chain now only requires $N/12+1$ qubits. 
% \sout{The DW2Q has been benchmarked against 
% another Ising physics\hyp{}based analog optimizer manufactured by Stanford University and NTT Corporation,
% and large performance differences are mostly attributed to the Chimera graph sparse connectivity.}

Going forward, we will benefit from QA technology improvements from the international community manufacturing quantum annealers with advanced capabilities. According to the development roadmap for these next\hyp{}generation quantum optimizers, it is expected that in \emph{ca.} a decade a system such as \systemname{} could be based on chips with tens of thousands of highly-connected qubits, with annealing schedules capable of more advanced quantum effects (\emph{e.g.} non\hyp{}stoquasticity~\cite{novikov2018exploring}) and engineering advances will have  order-of-magnitude improvements on the aforementioned overhead operation times. While quantum annealers are ahead in terms of number of qubits, gate-model systems offer additional controls that may conceivably increase performance in the future. We will investigate MIMO ML decoding on gate-model QPUs in future work, which currently cannot support algorithms that decode more than 4$\times$4 BPSK.

\section*{Acknowledgements}
%\parahead{ACKNOWLEDGEMENTS}
We thank our shepherd John Heidemann and
the anonymous reviewers for their insightful feedback. We thank the NASA Quantum AI Laboratory (QuAIL) and D-Wave Systems for useful discussions. The research is supported by National Science Foundation
%has received funding from National Science Foundation (NSF) 
(NSF) Awards \#1824357 and \#1824470 and by the USRA Cycle 3 Research Opportunity Program that allowed machine time on the DW2Q hosted at NASA Ames Research Center. Kyle Jamieson and Minsung Kim are partially supported by the Princeton University School of Engineering and Applied Science Innovation Fund. This work does not raise any ethical issues.

\clearpage
\bibliographystyle{ACM-Reference-Format}

\begin{raggedright}
\bibliography{reference}
\end{raggedright}

\appendix
\begin{small}
\noindent Appendices are supporting material that has not been peer reviewed.
\end{small}
\section{QUBO Forms}
\label{s:qubo_forms}

% For $N\times N$ MIMO communications, the Maximum Likelihood (ML) detection solves
% \begin{align}
% \arg\min_{\mathbf{v}} \lVert \mathbf{H\mathbf{v} - y} \rVert^2
% \end{align}
% where $H \in{R^{N \times N}}$ is complex MIMO channel, $\mathbf{v}$ $\in{R^N}$ is transmitter's symbol candidate, and $y \in{R^N}$ is received symbol. 
We demonstrate how to transform $2\times 2$ BPSK MIMO Maximum Likelihood (ML) detection into the QUBO form. ML detection solves Eq.~\ref{eqn:ml},
% \begin{align}
% \arg\min_{\mathbf{v}} \lVert \mathbf{H\mathbf{v} - y} \rVert^2
% \end{align}
where
\begin{small}
\begin{align}
\mathbf{H}=
\begin{bmatrix}
    h_{11}       & h_{12}   \\
    h_{21}       & h_{22}  \\
\end{bmatrix}
=
\begin{bmatrix}
    h_{I,11}       & h_{I,12}  \\
    h_{I,21}       & h_{I,22}  \\
\end{bmatrix}
+
j
\begin{bmatrix}
    h_{Q,11}       & h_{Q,12}  \\
    h_{Q,21}       & h_{Q,22}  \\
\end{bmatrix},
\notag\\
y=
\begin{bmatrix}
    y_{1}         \\
    y_{2}        \\
\end{bmatrix}
=
\begin{bmatrix}
    y_{I,1}       \\
    y_{I,2}       \\
\end{bmatrix}
+
j
\begin{bmatrix}
    y_{Q,1}        \\
    y_{Q,2}        \\
\end{bmatrix}\,\textrm{and}\,
\mathbf{v}=
\begin{bmatrix}
    v_{1}         \\
    v_{2}        \\
\end{bmatrix}.\notag
\end{align}
\end{small}
The norm expansion in Eq.~\ref{eqn:ml} can be expressed as
% \begin{equation}   
% \label{eqn:encoding_ex}
% \hat{\mathbf{s}} = \arg\min_{\mathbf{E(s_{1,2})} \in |\mathcal{O}|^{2}}
% \left\lVert
% \begin{bmatrix}
%     y_{1}          \\
%     y_{2}          \\
% \end{bmatrix} -
% \begin{bmatrix}
%     h_{11}       & h_{12}   \\
%     h_{21}       & h_{22}  \\
% \end{bmatrix}\begin{bmatrix}
%     v_1          \\
%     v_2          \\
% \end{bmatrix}  \right\rVert^2
% \end{equation}

% % \left\lVert
% % \begin{bmatrix}
% %     y_{1}          \\
% %     y_{2}          \\
% % \end{bmatrix} -
% % \begin{bmatrix}
% %     h_{11}       & h_{12}   \\
% %     h_{21}       & h_{22}  \\
% % \end{bmatrix}\begin{bmatrix}
% %     v_{1}          \\
% %     v_{2}          \\
% % \end{bmatrix}  \right\rVert^2
% % \leftrightarrow
\begin{small}
\begin{align}
\lVert \mathbf{y - Hv} \rVert^2=
\left\lVert
\begin{bmatrix}
    y_{1}          \\
    y_{2}          \\
\end{bmatrix} -
\begin{bmatrix}
    h_{11}       & h_{12}   \\
    h_{21}       & h_{22}  \\
\end{bmatrix}\begin{bmatrix}
    v_1          \\
    v_2          \\
\end{bmatrix}  \right\rVert^2 =
\begin{Vmatrix} 
y_1 - h_{11}v_1 - h_{12}v_2  \\
y_2 - h_{21}v_1 - h_{22}v_2  \\
\end{Vmatrix}^2\notag
\\
% =
% \begin{Vmatrix} 
% (h_{I,11}+ jh_{Q,11})v_1 + (h_{I,12}+ jh_{Q,12})v_2 - (y_{I,1} +jy_{Q,1}) \\
% (h_{I,21}+ jh_{Q,21})v_1 + (h_{I,22}+ jh_{Q,22})v_2 - (y_{I,2} +jy_{Q,2})
% \end{Vmatrix}^2 \notag\\
=\begin{Vmatrix} 
(y_{I,1} -h_{I,11}v_1 - h_{I,12}v_2) + j(y_{Q,1} -h_{Q,11}v_1 - h_{Q,12}v_2) \\
(y_{I,2} - h_{I,21}v_1 - h_{I,22}v_2) + j(y_{Q,2} - h_{Q,21}v_1 - h_{Q,22}v_2)
\end{Vmatrix}^2 \notag\\
= \{(y_{I,1} - h_{I,11}v_1 - h_{I,12}v_2)\}^2 + \{(y_{Q,1} - h_{Q,11}v_1 - h_{Q,12}v_2)\}^2 \notag\\
+ \{(y_{I,2} - h_{I,21}v_1 - h_{I,22}v_2)\}^2 + \{(y_{Q,2} - h_{Q,21}v_1 - h_{Q,22}v_2)\}^2.\notag
\end{align}
\end{small}
% The QUBO form is 
% \begin{align}
% \min_{q_i,q_j} \sum^{N}_{i\leq j=1} Q_{ij}q_i q_j
% \end{align}
% where Q $\in{R^{N \times N}}$ is an upper triangular matrix, and $q\in \{0,1\}$ is binary so $q=q^2$. In our problem, $q$ denotes qubit and $N$ the number of qubits. Note that i
In the case of BPSK, symbol $v_i \in \{-1,1\}$ is represented by a QUBO variable $q_i$. One possible transform is $2q_i-1$ where $q_i=0$ corresponds to $v_i=-1$ and $q_i=1$ to $v_i=1$. This leads to $\mathbf{v} = [v_{1}, v_{2}]^\intercal$ $=[\mathbf{T({q_1})}, \mathbf{T({q_2})}]^\intercal$, where $\mathbf{T(q_1)} = 2q_{1}-1$ and $\mathbf{T(q_2)} = 2q_{2}-1$. Using these relationships, we can express the above norm as
\begin{small}
\begin{align}
\lVert \mathbf{y - Hv} \rVert^2= \left\lVert
\begin{bmatrix}
    y_{1}          \\
    y_{2}          \\
\end{bmatrix} -
\begin{bmatrix}
    h_{11}       & h_{12}   \\
    h_{21}       & h_{22}  \\
\end{bmatrix}\begin{bmatrix}
    \mathbf{T(q_1)}          \\
    \mathbf{T(q_2)}          \\
\end{bmatrix}  \right\rVert^2 
\notag\\
=\{(y_{I,1} - h_{I,11}(2q_1-1) - h_{I,12}(2q_2-1))\}^2 \notag\\+  \{(y_{Q,1} - h_{Q,11}(2q_1-1) - h_{Q,12}(2q_2-1))\}^2  \notag\\+ \{( y_{I,2} - h_{I,21}(2q_1-1) - h_{I,22}(2q_2-1))\}^2 \notag\\+ \{(y_{Q,2} - h_{Q,21}(2q_1-1) - h_{Q,22}(2q_2-1))\}^2.\notag
\end{align}
\end{small}

Then
% \begin{small}
% \begin{align}
% \lVert \mathbf{y - Hv} \rVert^2= 
% \{2h_{I,11}^2 + 2(2h_{I,11})(-h_{I,11}-h_{I,12}-y_{I,1})\notag\\
% + 2h_{Q,11}^2 + 2(2h_{Q,11})(-h_{Q,11}-h_{Q,12}-y_{Q,1})\notag\\
% + 2h_{I,21}^2 + 2(2h_{I,21})(-h_{I,21}-h_{I,22}-y_{I,2})\notag\\
% + 2h_{Q,21}^2 + 2(2h_{Q,21})(-h_{Q,21}-h_{Q,22}-y_{Q,2})\}q_1\notag\\ 
% +\{2h_{I,12}^2 + 2(2h_{I,12})(-h_{I,11}-h_{I,12}-y_{I,1})\notag\\
% + 2h_{Q,12}^2 + 2(2h_{Q,12})(-h_{Q,11}-h_{Q,12}-y_{Q,1})\notag\\
% + 2h_{I,22}^2 + 2(2h_{I,22})(-h_{I,21}-h_{I,22}-y_{I,2})\notag\\
% + 2h_{Q,22}^2 + 2(2h_{Q,22})(-h_{Q,21}-h_{Q,22}-y_{Q,2})\}q_2\notag\\
% +\{2(2h_{I,11})(2h_{I,12})+2(h_{Q,11})(h_{Q,12})\notag\\
% + 2(2h_{I,21})(2h_{I,22})+2(h_{Q,21})(h_{Q,22})\}q_1q_2\notag\\
% + \{(-h_{I,11}-h_{I,12}-y_{I,1})^2 + (-h_{Q,11}-h_{Q,12}-y_{Q,1})^2\notag\\
% + (-h_{I,21}-h_{I,22}-y_{I,2})^2 + (-h_{Q,21}-h_{Q,22}-y_{Q,2})^2\}.
% \end{align}
% \end{small}
we obtain the objective function of ML problem with QUBO variables. Using $q_i^2=q_i$, minimization of this objective function becomes the QUBO form (Eq.~\ref{eqn:qubo}): 
\begin{small}
\begin{align}
\hat{q_1}, \hat{q_2} = \arg\min_{q_1,q_2} Q_{11}q_1 + Q_{22}q_2 + Q_{12}q_1q_2, \,\textrm{where}\,\notag\\
% \min_{q_i,q_j} \sum^{2}_{i\leq j=1} Q_{ij}q_i q_j
% , \textrm{where} \, Q = 
% \begin{bmatrix}
%     Q_{11} & Q_{12}         \\
%     0     & Q_{22}       \\
% \end{bmatrix},\notag \\
Q_{11} = -4h_{I,11}y_{I,1} -4h_{I,21}y_{I,2} -4h_{Q,11}y_{Q,1} -4h_{Q,21}y_{Q,2} \notag\\-4h_{I,11}h_{I,12} -4h_{I,21}h_{I,22} -4h_{Q,11}h_{Q,12}- 4h_{Q,21}h_{Q,22},\notag\\
Q_{22} = -4h_{I,12}y_{I,1} -4h_{I,22}y_{I,2} -4h_{Q,12}y_{Q,1} -4h_{Q,22}y_{Q,2} \notag\\-4h_{I,12}h_{I,12} -4h_{I,22}h_{I,22} -4h_{Q,12}h_{Q,12}- 4h_{Q,22}h_{Q,22},\notag\\
Q_{12} = 8h_{I,11}h_{I,12} + 8h_{I,21}h_{I,22} + 8h_{Q,11}h_{Q,12} + 8h_{Q,21}h_{Q,22}.\notag
% Q_{11} = 2h_{I,11}^2 + 2(2h_{I,11})(-h_{I,11}-h_{I,12}-y_{I,1})\notag\\
% + 2h_{Q,11}^2 + 2(2h_{Q,11})(-h_{Q,11}-h_{Q,12}-y_{Q,1})\notag\\
% + 2h_{I,21}^2 + 2(2h_{I,21})(-h_{I,21}-h_{I,22}-y_{I,2})\notag\\
% + 2h_{Q,21}^2 + 2(2h_{Q,21})(-h_{Q,21}-h_{Q,22}-y_{Q,2}),\notag\\
% Q_{22} = 2h_{I,12}^2 + 2(2h_{I,12})(-h_{I,11}-h_{I,12}-y_{I,1})\notag\\
% + 2h_{Q,12}^2 + 2(2h_{Q,12})(-h_{Q,11}-h_{Q,12}-y_{Q,1})\notag\\
% + 2h_{I,22}^2 + 2(2h_{I,22})(-h_{I,21}-h_{I,22}-y_{I,2})\notag\\
% + 2h_{Q,22}^2 + 2(2h_{Q,22})(-h_{Q,21}-h_{Q,22}-y_{Q,2}),\notag\\
% Q_{12} = 2(2h_{I,11})(2h_{I,12})+2(h_{Q,11})(h_{Q,12})\notag\\
% + 2(2h_{I,21})(2h_{I,22})+2(h_{Q,21})(h_{Q,22}).
\end{align}
\end{small}

\section{Embedded Ising}
\label{s:emb-ising}
Embedding maps the Ising problem to an equivalent one
that has the same ground state, but also satisfies Chimera graph 
constraints. 
The \systemname{} compiled objective function is:
\begin{eqnarray} 
-\sum_{i=1}^N\left[ \sum_{c=1}^{\lceil N/4 \rceil} s_{ic} s_{i(c+1)}\right]&&\label{eq:IsingEmbeddedObjFunJF}\\
+\sum_{i=1}^{N} \left(\frac{\bf{f_i}}{|J_F|\left(\lceil\frac{N}{4}\rceil +1\right)}\right) \left[\sum_{c=1}^{\lceil N/4\rceil+1 } s_{ic}\right]&&\label{eq:IsingEmbeddedObjFun1}\\
+ \sum_{i,j=1}^{N} \frac{\bf{g_{ij}}}{|J_F|} \sum_{(c_i,c_j)\in\delta_{ij}} s_{ic_i}s_{jc_j}&&\label{eq:IsingEmbeddedObjFun2}
\end{eqnarray}
where the original logical variables $s_i$ are now associated to a chain $i$
of $c=1\dots (\lceil N/4 \rceil+1)$ qubits, indexed with new spins $s_{ic}$. 
$|J_F|$ 
penalizes the condition that $s_{ic}\neq s_{ic^\prime}$, {\it i.e.}, enforces that
all qubits in the chain assume the same value ($\pm 1$). This enforcement is more likely
to happen for large values of $|J_F|$, however the maximum negative energy value is 
set to $-1$ by hardware design. In (\ref{eq:IsingEmbeddedObjFun1}) and (\ref{eq:IsingEmbeddedObjFun2}), $|J_F|$ effectively renormalizes 
all terms in the objective function by the factor $|J_F|^{-1}$.
The linear term value $\bf{f_i}$ is additionally divided by the number of qubits
in a chain ($\lceil N/4 \rceil+1$). The term in (\ref{eq:IsingEmbeddedObjFun2}) shows that
the duplication of variables ensures the existence of a pair of qubits in the chains
such that a physical coupler in the Chimera graph exists ($\delta_{ij}$ is the
set of pairs of qubits that are connected by a physical bond once the chains $i$ and $j$
are specified).
% \begin{eqnarray} 
% -\sum_{i=1}^N\left[ \sum_{c=1}^{N/4} s_{ic}^z s_{i(c+1)}^z\right]&&\label{eq:IsingEmbeddedObjFunJF}\\
% +\sum_{i=1}^{N} \left(\frac{\bf{f_i}}{|J_F|\left(\frac{N}{4}+1\right)}\right) \left[\sum_{c=1}^{N/4+1} s_{ic}\right]&&\label{eq:IsingEmbeddedObjFun1}\\
% + \sum_{i,j=1}^{N} \frac{\bf{g_{ij}}}{|J_F|} \sum_{(c_i,c_j)\in\delta_{ij}} s_{ic_i}s_{jc_j}&&\label{eq:IsingEmbeddedObjFun2}
% \end{eqnarray}
\section{16\hyp{}QAM Ising Model Parameters}
\label{s:16qm_model_parameter}

Following are the Ising parameters $f_{i}$ for 16\hyp{}QAM:

\begin{small}
\begin{align}
    f_{i}(\mathbf{H}, \mathbf{y}) = 
\begin{cases}
    \text{case } i = 4n-3:\\
     -4\left(\mathbf{H}^{I}_{(:,\lceil i/4\rceil)}\cdot  \mathbf{y}^{I}\right) -4 \left(\mathbf{H}^{Q}_{(:,\lceil i/4\rceil)}\cdot \mathbf{y}^{Q}\right),\\
     \text{case } i = 4n-2:\\
     -2\left(\mathbf{H}^{I}_{(:,\lceil i/4\rceil )}\cdot  \mathbf{y}^{I}\right) -2 \left(\mathbf{H}^{Q}_{(:,\lceil i/4\rceil)}\cdot \mathbf{y}^{Q}\right),\\      \text{case } i = 4n-1:\\
     -4\left(\mathbf{H}^{I}_{(:,\lceil i/4\rceil)}\cdot  \mathbf{y}^{Q}\right) +4 \left(\mathbf{H}^{Q}_{(:,\lceil i/4\rceil)}\cdot \mathbf{y}^{I}\right),\\
     \text{case } i = 4n:\\
     -2\left(\mathbf{H}^{I}_{(:,\lceil i/4\rceil )}\cdot  \mathbf{y}^{Q}\right) +2 \left(\mathbf{H}^{Q}_{(:,\lceil i/4\rceil)}\cdot \mathbf{y}^{I}\right).
\end{cases}
\label{eqn:16qam_ising_f}
\end{align} 
\end{small}

Since real and imaginary terms of each symbol are independent, the coupler 
strength between $s_{4n-3}$, $s_{4n-2}$ and $s_{4n-1}$, $s_{4n}$ is 0.
For other $s_i$ and $s_j$, the Ising coupler strength
$g_{ij}$ for 16\hyp{}QAM is:

\begin{small}
\begin{align}
    g_{ij}(\mathbf{H}) = 
\begin{cases}
    \text{case } i = 4n-3:\\  %%
    \begin{cases}
    \text{case } j = 4n'-3:\\
     8\left(\mathbf{H}^{I}_{(:,\lceil i/4\rceil)}\cdot \mathbf{H}^{I}_{(:,\lceil j/4\rceil)}\right)+8\left(\mathbf{H}^{Q}_{(:,\lceil j/4\rceil)}\cdot \mathbf{H}^{Q}_{(:,\lceil i/4\rceil)}\right),\\
    \text{case } j = 4n'-2:\\
    4\left(\mathbf{H}^{I}_{(:,\lceil i/4\rceil)}\cdot \mathbf{H}^{I}_{(:,\lceil j/4\rceil)}\right)+4\left(\mathbf{H}^{Q}_{(:,\lceil j/4\rceil)}\cdot \mathbf{H}^{Q}_{(:,\lceil i/4\rceil)}\right),\\
    \text{case } j = 4n'-1:\\
    -8\left(\mathbf{H}^{I}_{(:,\lceil i/4\rceil)}\cdot \mathbf{H}^{Q}_{(:,\lceil j/4\rceil)}\right)+8\left(\mathbf{H}^{I}_{(:,\lceil j/4\rceil)}\cdot \mathbf{H}^{Q}_{(:,\lceil i/4\rceil)}\right),\\
    \text{case } j = 4n':\\
    -4\left(\mathbf{H}^{I}_{(:,\lceil i/4\rceil)}\cdot \mathbf{H}^{Q}_{(:,\lceil j/4\rceil)}\right)+4\left(\mathbf{H}^{I}_{(:,\lceil j/4\rceil)}\cdot \mathbf{H}^{Q}_{(:,\lceil i/4\rceil)}\right),
    \end{cases}\\
    %split?
    \text{case } i = 4n-2:\\  %%
    \begin{cases}
    \text{case } j = 4n'-3:\\
     4\left(\mathbf{H}^{I}_{(:,\lceil i/4\rceil)}\cdot \mathbf{H}^{I}_{(:,\lceil j/4\rceil)}\right)+4\left(\mathbf{H}^{Q}_{(:,\lceil j/4\rceil)}\cdot \mathbf{H}^{Q}_{(:,\lceil i/4\rceil)}\right),\\
    \text{case } j = 4n'-2:\\
    2\left(\mathbf{H}^{I}_{(:,\lceil i/4\rceil)}\cdot \mathbf{H}^{I}_{(:,\lceil j/4\rceil)}\right)+2\left(\mathbf{H}^{Q}_{(:,\lceil j/4\rceil)}\cdot \mathbf{H}^{Q}_{(:,\lceil i/4\rceil)}\right),\\
    \text{case } j = 4n'-1:\\
    -4\left(\mathbf{H}^{I}_{(:,\lceil i/4\rceil)}\cdot \mathbf{H}^{Q}_{(:,\lceil j/4\rceil)}\right)+4\left(\mathbf{H}^{I}_{(:,\lceil j/4\rceil)}\cdot \mathbf{H}^{Q}_{(:,\lceil i/4\rceil)}\right),\\
    \text{case } j = 4n':\\
    -2\left(\mathbf{H}^{I}_{(:,\lceil i/4\rceil)}\cdot \mathbf{H}^{Q}_{(:,\lceil j/4\rceil)}\right)+2\left(\mathbf{H}^{I}_{(:,\lceil j/4\rceil)}\cdot \mathbf{H}^{Q}_{(:,\lceil i/4\rceil)}\right),
    \end{cases}\\
    %
    %%split
    \text{case } i = 4n-1:\\ %%
    \begin{cases}
    \text{case } j = 4n'-3:\\ 
     8\left(\mathbf{H}^{I}_{(:,\lceil i/4\rceil)}\cdot \mathbf{H}^{Q}_{(:,\lceil j/4\rceil)}\right)-8\left(\mathbf{H}^{I}_{(:,\lceil j/4\rceil)}\cdot \mathbf{H}^{Q}_{(:,\lceil i/4\rceil)}\right),\\
    \text{case } j = 4n'-2:\\
    4\left(\mathbf{H}^{I}_{(:,\lceil i/4\rceil)}\cdot \mathbf{H}^{Q}_{(:,\lceil j/4\rceil)}\right)-4\left(\mathbf{H}^{I}_{(:,\lceil j/4\rceil)}\cdot \mathbf{H}^{Q}_{(:,\lceil i/4\rceil)}\right),\\
    \text{case } j = 4n'-1:\\
    8\left(\mathbf{H}^{I}_{(:,\lceil i/4\rceil)}\cdot \mathbf{H}^{I}_{(:,\lceil j/4\rceil)}\right)+8\left(\mathbf{H}^{Q}_{(:,\lceil j/4\rceil)}\cdot \mathbf{H}^{Q}_{(:,\lceil i/4\rceil)}\right),\\
    \text{case } j = 4n':\\
    4\left(\mathbf{H}^{I}_{(:,\lceil i/4\rceil)}\cdot \mathbf{H}^{I}_{(:,\lceil j/4\rceil)}\right)+4\left(\mathbf{H}^{Q}_{(:,\lceil j/4\rceil)}\cdot \mathbf{H}^{Q}_{(:,\lceil i/4\rceil)}\right),
    \end{cases}\\
    \text{case } i = 4n:\\ %%
    \begin{cases}
    \text{case } j = 4n'-3:\\ 
     4\left(\mathbf{H}^{I}_{(:,\lceil i/4\rceil)}\cdot \mathbf{H}^{Q}_{(:,\lceil j/4\rceil)}\right)-4\left(\mathbf{H}^{I}_{(:,\lceil j/4\rceil)}\cdot \mathbf{H}^{Q}_{(:,\lceil i/4\rceil)}\right),\\
    \text{case } j = 4n'-2:\\
    2\left(\mathbf{H}^{I}_{(:,\lceil i/4\rceil)}\cdot \mathbf{H}^{Q}_{(:,\lceil j/4\rceil)}\right)-4\left(\mathbf{H}^{I}_{(:,\lceil j/4\rceil)}\cdot \mathbf{H}^{Q}_{(:,\lceil i/4\rceil)}\right),\\
    \text{case } j = 4n'-1:\\
    4\left(\mathbf{H}^{I}_{(:,\lceil i/4\rceil)}\cdot \mathbf{H}^{I}_{(:,\lceil j/4\rceil)}\right)+4\left(\mathbf{H}^{Q}_{(:,\lceil j/4\rceil)}\cdot \mathbf{H}^{Q}_{(:,\lceil i/4\rceil)}\right),\\
    \text{case } j = 4n':\\
    2\left(\mathbf{H}^{I}_{(:,\lceil i/4\rceil)}\cdot \mathbf{H}^{I}_{(:,\lceil j/4\rceil)}\right)+2\left(\mathbf{H}^{Q}_{(:,\lceil j/4\rceil)}\cdot \mathbf{H}^{Q}_{(:,\lceil i/4\rceil)}\right).
    \end{cases}\\
\end{cases}
%\notag
%\label{eqn:16qam_ising_g}
\end{align} 
\end{small}

\end{document}